\documentclass[12pt]{article}

\usepackage{epsf,epsfig}

\usepackage{a4wide}

 \newlength{\dinwidth}
 \newlength{\dinmargin}
\def\beq{\begin{eqnarray}}
\def\eeq{\end{eqnarray}}

\def\be{\begin{equation}}
\def\ee{\end{equation}}

\def\slash#1{#1 \hskip-0.50em /}
\def\Slash#1{#1 \hskip-0.65em /}

\begin{document}

 \newcommand{\bea}{\begin{eqnarray}}
 \newcommand{\eea}{\end{eqnarray}}
 \newcommand{\nn}{\nonumber}
 \newcommand{\dd}{\displaystyle}
 \newcommand{\bra}[1]{\left\langle #1 \right|}
 \newcommand{\ket}[1]{\left| #1 \right\rangle}
 \newcommand{\spur}[1]{\not\! #1 \,}

\thispagestyle{empty}

\begin{flushright}
  CERN-PH-TH/2005-053\\
  SLAC-PUB-11082\\
  SI-HEP-2005-03\\
  BARI-TH/05-509\\
  hep-ph/0504088
\end{flushright}

\vspace{\baselineskip}

\begin{center}

{\bf \Large
Light-cone sum rules\\[0.5em] in
soft-collinear effective theory}

\vspace{3.5\baselineskip}
{\large
  Fulvia De Fazio$^a$,
  Thorsten Feldmann$^{b}$,
  and
  Tobias Hurth$^{c,d,}$\footnote{Heisenberg Fellow}}
\\
\vspace{2em}

{  \it
$^a$ Istituto Nazionale di Fisica Nucleare, Sezione di Bari, Italy}
\\[0.3em]
{  \it
$^b$ Fachbereich Physik,
  Universit\"at Siegen, D-57068 Siegen, Germany}
\\[0.3em]
{  \it
$^c$ CERN, Dept.\ of Physics, Theory Division, CH-1211 Geneva 23, Switzerland}
\\[0.3em]
{ \it $^d$ SLAC, Stanford University, Stanford, CA 94309, USA}

\vspace{3\baselineskip}

%
\textbf{Abstract}
\vspace{1em}

\parbox{0.9\textwidth}
{
We derive light-cone sum rules (LCSRs) for exclusive $B$\/-meson decays
into light energetic hadrons from correlation functions
within soft-collinear effective theory (SCET). In these sum rules the
short-distance scale refers to ``hard-collinear'' interactions with
virtualities of order $\Lambda_{\rm QCD} m_b$. Hard scales (related
to virtualities of order $m_b^2$) are integrated out and enter via
external coefficient functions in the sum rule. Soft dynamics is
encoded in light-cone distribution amplitudes for the
$B$\/-meson, which describe both the factorizable and non-factorizable
contributions to exclusive $B$\/-meson decay amplitudes.
As an example, we provide a detailed study of the SCET sum rule
for the $B \to \pi$ transition form factor at large recoil, including
radiative corrections from hard-collinear loop diagrams
at first order in the strong coupling constant.
We find remarkable conceptual and numerical
differences with the heavy-quark limit of the
conventional LCSR approach in QCD.
}

\end{center}

\clearpage
\setcounter{page}{1}

\section{Introduction}
\label{sec:intro}

$B$-meson decays to a pseudoscalar ($P$) or a vector meson ($V$) involve
(among others) hadronic matrix elements that define $B \to P$ and
$B \to V$ transition form factors. Three form factors are
required to describe $B \to P$ transitions, while seven are needed
in the $B \to V$ case. These form factors represent an important source
of hadronic uncertainties to the determination of the CKM element
$|V_{ub}|$ from exclusive semi-leptonic $b \to u$ decays
\cite{Athar:2003yg,Abe:2004zm,Aubert:2004bq} or to the
extraction of the CKM angle $\alpha$ from charmless non-leptonic
$B$\/-decays in the QCD factorization approach \cite{Beneke:1999br}.
Let us consider the case in which
the meson in the final state  is a light one. Near zero momentum
transfer ($q^2 \simeq 0$) the flavour-changing weak current
transforms a $B$~meson in its rest frame into a highly energetic
hadron. The transition form factor reflects the internal dynamics
that distributes the large energy release among the constituents
of the final-state hadron.
The energy scales involved in these processes are: i)
$\Lambda = {\rm few} \times \Lambda_{\rm QCD}$,
the {\it soft} scale set by the typical energies and
momenta of the light degrees of freedom in the hadronic bound states.
ii) $m_b$ the {\it hard}\/ scale set by the heavy-$b$\/-quark mass.
Notice that in the $B$\/-meson rest frame for $q^2\simeq 0$
also the energy of the final-state hadron is given by $E\simeq m_b/2$.
iii) The hard-collinear scale $\mu_{\rm hc}=\sqrt{m_b
\Lambda}$ appears via interactions between soft and energetic
modes in the initial and final state. The dynamics of hard and hard-collinear
modes can be described perturbatively in the heavy-quark limit 
$m_b \to \infty$.
The separation of the two perturbative scales 
from the non-perturbative hadronic
dynamics is formalized within the framework of soft-collinear effective theory
(SCET) \cite{Bauer:2000yr,Beneke:2002ph}. The small expansion parameter
in SCET is given by the ratio $\lambda = \sqrt{\Lambda/m_b}$, in terms of
which the hierarchy between the hard scale,
the hard-collinear scale, and the soft scale reads
\beq
&&  \mu_0 \sim \lambda^2 m_b \ll \mu_{\rm hc} \sim \lambda m_b \ll
  \mu_{\rm h} \sim m_b \, .
\eeq

A detailed study of the heavy-to-light transition form factors
\cite{Beneke:2003pa} (see also \cite{Lange:2003pk,Bauer:2002aj})
shows that in the heavy-quark limit each form factor can
be decomposed into two basic contributions:
\begin{itemize}
  \item One contribution factorizes into a perturbatively calculable
        coefficient function and light-cone distribution amplitudes
        $\phi_B$ and $\phi_\pi$
        for heavy and light mesons, respectively.
        The former describes the short-distance
        interactions between the decay current and the spectator quarks
        by hard-collinear gluon exchange and is proportional
        to $\alpha_s(\mu_{\rm hc})$.
        The latter can be considered as probability
        amplitudes to find a quark-antiqurak pair
        with certain light-cone momentum
        fractions inside the hadron.
  \item In the second contribution, the hard-collinear interactions are
        {\em not}\/ factorizable, leaving one universal ``soft'' form factor
        for each type of meson (pseudoscalar, longitudinally or transversely
        polarized vector, etc.); this  does not depend on the Dirac structure
        of the decay current.
        Because of the non-factorizable nature, the ``soft''
        form factor is in general a non-perturbative object
        of order $(\alpha_s)^0$.
        A still controversial issue is the question
        as to what extent it is numerically
        suppressed by Sudakov effects
        (see for instance \cite{Descotes-Genon:2001hm}
        for a critical discussion).
\end{itemize}

In the following we restrict ourselves to the pion case
(the generalization to other mesons should be obvious).
Schematically the decomposition of the transition form factors  reads
\cite{Beneke:2000wa}
\be
\langle \pi| \bar \psi \, \Gamma_i \, b |B\rangle
=  C_i(E, \mu_I) \, \xi_\pi(\mu_I,E) +
   T_i(E,u,\omega,\mu_{\rm II}) \otimes \phi_+^B(\omega,\mu_{\rm II}) \otimes
         \phi_\pi(u,\mu_{\rm II})+
   \ldots,
\label{factorization}
\ee
where the dots  stand for sub-leading terms in $\Lambda/m_b$.
Here $C_i$ is a short-distance function arising from integrating out
hard modes, and consequently $\mu_I$ is a factorization scale
below $m_b$;
$T_i$ is the hard-scattering function mentioned above, which contains
the effect of both hard and hard-collinear dynamics,  $\mu_{\rm II}$ being
a factorization scale below $\mu_{\rm hc}$.
Both functions can be computed as perturbative series in $\alpha_s$, and
potentially large logarithms $\ln m_b/\mu_{\rm I}$ and $\ln \mu_{\rm hc}/\mu_{\rm II}$
can be resummed by (more or less)
standard renormalization-group techniques
(the effective theories for the two short-distance regimes are known as
SCET$_{\rm I}$ and SCET$_{\rm II}$, respectively).

For instance, using the notation of \cite{Bauer:2000yr},
the vector current in QCD with $\Gamma_i=\gamma_\mu$
in (\ref{factorization})
is matched onto
\beq
  \bar q \, \gamma_\mu \, b & \longrightarrow &
  \left( C_4 n_-^\mu +C_5 v^\mu \right) {\bar \xi}_{\rm hc} W_{\rm hc} \,  Y_s^\dagger  h_v
 + \mbox{sub-leading terms}
\label{matching0}
\eeq
in SCET$_{\rm I}$ with $C_4=1 + {\cal O}(\alpha_s)$
and $C_5 = {\cal O}(\alpha_s)$. Here we have introduced the
heavy-quark velocity $v^\mu$ and
have chosen two light-like vectors $n_+^\mu$ and
$n_-^\mu$ which are normalized as $n_+ n_- =2$ and $n_\pm v = 1$.
The direction of the momentum of the (massless) pion is given by
$p_\pi^\mu = (n_+p_\pi) \, n_-^\mu/2$.
Neglecting order $\alpha_s$ effects one
obtains in this way approximate relations between
the vector and tensor form factors for $B \to \pi$ transitions
\cite{Charles:1998dr,Beneke:2000wa}:
$$
  f_+(q^2) \simeq \frac{m_B}{n_+p_\pi} \, f_0(q^2)
  \simeq \frac{m_B}{m_B+m_\pi} \, f_T(q^2)
  \simeq \xi_\pi(q^2) \, .
$$
The soft form factor entering (\ref{factorization}) can be defined as
\cite{Beneke:2003pa}
\beq
  \langle \pi(p') |
  (\bar \xi_{\rm hc} W_{\rm hc})(0) \, (Y_s^\dagger h_v)(0) |B(m_B v)\rangle
  &=&
  (n_+p') \, \xi_\pi(n_+p', \, \mu_{\rm I}) \,,
\label{eq:softdef}
\eeq
where
\beq
\xi_{\rm hc}(x) &=& \frac{\slash n_- \slash n_+}{4} \, \psi_{\rm hc}(x)
\eeq
is a hard-collinear light-quark field in SCET$_{\rm I}$, and
\beq
h_v(x) = \frac{1+\slash v}{2} \, e^{i m_b v x} \, b(x)
\eeq
is an HQET field \cite{Neubert:1993mb}.
The hard-collinear and soft Wilson lines $W_{\rm hc}$ and $Y_s$ appear to
render the definition gauge-invariant;
for their definition, see for instance \cite{Beneke:2003pa}.

The hard-scattering functions $T_i$ only enter at ${\cal O}(\alpha_s)$.
They involve the LCDAs of the $B$ meson
\cite{Grozin:1996pq}
and of the pion \cite{Braun:1989iv},
which are defined as
\beq
  \langle 0|\bar \psi(x) \, P(x,0) \, \slash x \, \gamma_5  
      \, h_v(0)|B(m_B v)\rangle &=&
  i f_B(\mu) \, m_B (v \cdot x) \,
   \int_0^\infty d\omega \, e^{-i\omega \, v \cdot x} \, \phi_+^B(\omega,\mu)
\\[0.3em]
\langle \pi(p')|\bar \psi(y) \, P(y,0) \, \slash y \, \gamma_5 \,
               \psi(0)|0\rangle   &=&
- i \, f_\pi \, (p' \cdot y) \, \int_0^1 du \, e^{i u \, p' \cdot y} \, \phi_\pi(u,\mu)\,,
\eeq
with $x^2=y^2=0$ and $P(x,0)$ being a gauge-link factor
along a straight path between $x$ and $0$. In the SCET framework the fields
in the above definitions are restricted to soft modes for the $B$ meson and
collinear modes for the pion, and one may choose $x^\mu \parallel n_-^\mu$ and
$y^\mu \parallel n_+^\mu$. Then $\omega = n_- p_q$
can be interpreted as the light-cone
momentum of the soft spectator quark in the $B$ meson, whereas $u = n_+p_q'/n_+p'$
is the light-cone momentum fraction of a quark inside the pion.

SCET thus provides a field-theoretical framework to achieve the factorization
of short- and long-distance physics, and to calculate the former in
renormalization-group-improved perturbation theory. However, the non-perturbative
dynamics encoded in the light-cone distribution amplitudes and the soft form
factors remains undetermined without further phenomenological or theoretical input.
In the past, the two ``standard'' tools to deal with non-perturbative dynamics in
QCD were  space-time lattice simulations
(see for instance \cite{lattice1,lattice2,lattice3}) and QCD/light-cone sum rules
(see for instance \cite{Khodjamirian,Ball,Colangelo:2000dp}).
The specific features of the heavy-quark expansion in
heavy-to-light transitions
at large recoil are reflected in the rather
complicated dynamics encoded in SCET.
In particular, the non-local nature of the SCET Lagrangian and decay currents
prevents (for the moment) a direct computation of SCET matrix elements on
the lattice; the lattice
calculation of $B \to \pi$ form factors in QCD is restricted
to the kinematical range
where the energy transfer to the pion is not too large
(see however \cite{Foley:2002qv} for a recent development to overcome
 this restriction).
On the other hand, light-cone sum rules seem to be very closely related to the
SCET formulation; see \cite{Ball:2003bf} for a detailed comparison.
Nevertheless,
we will show in this article that sum rules that can be formulated for the {\em soft}\/ (i.e.\ non-factorizable)
part of the form factor {\em within SCET}\/ are different from those in the conventional light-cone
sum-rule approach.
The difference arises as follows:
In the conventional approach one starts from
an appropriate correlation function, where the heavy-$B$ meson is
replaced by a heavy-light current, and the light-meson state
is expanded in terms of light-cone wave functions of increasing twist.
The dispersive analysis of the
correlation function with the usual sum-rule techniques is
performed for finite
heavy-quark masses. Only at the very end of the calculation may
the heavy-quark limit
be taken and factorizable and non-factorizable contributions
be identified.

In contrast, within SCET the heavy-quark expansion is performed
at the very beginning of every calculation.
The light-cone separation of composite operators {\em follows}\/
from the Feynman rules in SCET,
which also determine which of the hadrons should
be represented by an interpolating current and
which could be described in terms of
light-cone wave functions.
In particular, the resummation of Sudakov logarithms, which become
large in the heavy-mass limit, is under control in the effective theory.

The paper is organized as follows:
in the next Section we will critically review the traditional
light-cone sum rule for the $B \to \pi$ form factor
in the heavy-quark limit, and explain the
apparently
bad convergence of the heavy-quark expansion for the soft form factor
in that approach. In Section~3 we formulate the SCET sum rule, starting
from the tree-level expression for the soft form factor.
Radiative corrections to the soft form factor from hard-collinear
loop diagrams are shown to match the large Sudakov double logarithms
from the hard matching coefficients for decay currents in SCET
and the evolution of the $B$\/-meson distribution amplitude in HQET.
We also show that the sum rule for the factorizable decay current
reproduces the result from QCD factorization.
Section~4 is devoted to a numerical discussion of the
sum rule (in the so-called Wandzura-Wilczek approximation),
with particular emphasis on theoretical uncertainties.
A brief summary and outlook is presented in
Section~5.

\section{Conventional LSCR for
$B \to \pi$ form factor\\ (in the heavy-quark limit)}

\label{sec:conv}

In the traditional LCSR approach to exclusive $B$\/ decays\footnote{
Several applications
which often include also a
detailed discussion of the heavy
quark limit can be found for example in
\cite{Bagan,Balletal,Khodjamirianetal}.},
the $B$ meson is represented by an interpolating current
 $J_B$, and the sum rule is derived from the
QCD-correlation function between $J_B$ and a
weak decay operator $J_i=\bar q \, \Gamma_i \, b$.
This leads to matrix elements of some non-local
operator $\langle \pi | {\cal O}(x_1,x_2,\ldots) |0\rangle$.
The operator ${\cal O}$ is expanded on the light-cone,
and the hadronic matrix elements of this expansion define
light-cone distribution amplitudes (LCDAs) for the pion of increasing twist.

A subtle point in exclusive $B$ decays is related to end-point configurations,
where one of the partons carries almost all of the
energy of a light hadron in the final state. In the conventional LCSR approach,
these configurations, and the resulting ``soft'' contribution to the exclusive
decay amplitudes, can be traced back to terms that are sensitive to
LCDAs at the end-point (more precisely, to the first non-vanishing
term of a Taylor expansion around the end-point) \cite{Bagan}.
For example, the twist-2 and twist-3 contributions from two-particle
LCDAs to the form factor $f_+$ in $B \to \pi \ell\nu$ transitions
read (in the heavy-quark limit,
see Eqs.~(28) and (31) in \cite{Ball:2003bf}):\footnote{For 
illustrative purposes, we have referred to the so-called
finite-energy or local-duality limit of the sum rules, where
the Borel parameter is set to infinity. The dependence of the
soft contribution on $\phi_\pi'(1)$ and $\phi_P(1)$ in the
heavy-quark limit remains true beyond this approximation.}

\beq
  f_B m_b f_+^{T2}(0) &=& - f_\pi \frac{\omega_0^2}{m_b} \,
  \phi'_\pi(1,\mu) \left\{
   1 + \frac{\alpha_s C_F}{4\pi}
  \left( 1 + \pi^2 - 2 \ln^2 \frac{2\omega_0}{m_b} -
  4 \ln \frac{2 \omega_0}{m_b} + 2 \ln \frac{2\omega_0}{\mu}\right)
 \right\} \nonumber
\\[0.2em]
  && + 4 f_\pi \, \frac{\alpha_s C_F}{4\pi} \frac{\omega_0^2}{m_b}
 \left\{ \left( \ln \frac{2\omega_0}{\mu} - 1 \right)
  \int_0^1 du \frac{\phi_\pi(u) + \bar u \phi_\pi'(1)}{\bar u^2}
 + \ln \frac{2\omega_0}{\mu} \int_0^1 du \frac{\phi_\pi(u)}{\bar u}
\right\} \, ,
\label{soft2}
\cr &&
\\[0.2em]
f_B m_b f_+^{T3}(0) &=& 2 \mu_\pi^2(\mu) \frac{\omega_0}{m_b}
  \phi_P(1,\mu) \left\{ 1 + \frac{\alpha_s C_F}{4\pi}
\left( \pi^2 - 7 - 2 \ln^2 \frac{2\omega_0}{m_b}
  - 4 \ln \frac{2 \omega_0}{m_b} + 6 \ln \frac{2\omega_0}{\mu}\right)
\right\}
\nonumber \\[0.2em]
&& + 2 \mu_\pi^2(\mu) \frac{\omega_0}{m_b} \frac{\alpha_s C_F}{4\pi}
\left(4 \ln \frac{2\omega_0}{\mu} -3 \right)
\int_0^1 du \frac{\phi_P(u) - \phi_P(1)}{\bar u}
\nonumber \\[0.2em]
&& - \mu_\pi^2(\mu) \frac{\omega_0}{3m_b} \frac{\alpha_s C_F}{4\pi}
\int_0^1 du \frac{\phi_\sigma(u) + \bar u \phi_\sigma'(1)}{\bar u^2}\,,
\label{soft3}
\eeq
where $\phi_i(u,\mu)$ are LCDAs, $\omega_0 \sim \Lambda$
is related to the threshold parameter and $\mu_\pi
= m_\pi^2/2m_q$.

\begin{figure}[tbhp]
\begin{center}
\psfig{file=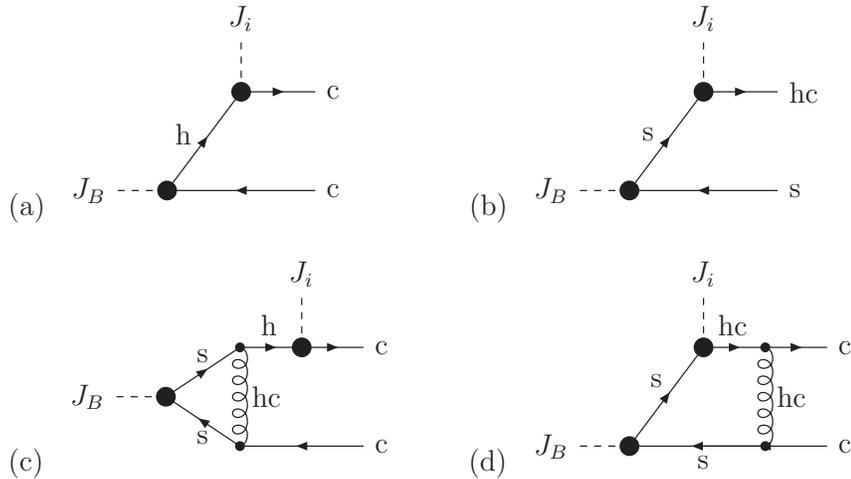, width=0.7\textwidth}
\end{center}
\caption{
Examples for different momentum configurations that contribute to 
the $B \to \pi$ form factor in the conventional LCSR
approach (``h'' stands for hard, ``hc'' for hard-collinear, ``c'' for
collinear, and ``s'' for soft
interactions, using the terminology of \cite{Beneke:2003pa}).
Notice that hard interactions refer to scales of order $m_b$,
hard-collinear interactions to scales of order $\sqrt{m_b \Lambda}$,
and soft and collinear interactions to scales of order $\Lambda$.
}
\label{plot1}
\end{figure}

In the following, we are critically re-examining the assumptions that are the
basis of the LCSR  approach to exclusive $B$ decays. As the
simplest example we consider the $B \to\pi$ form factor at large recoil.
Representing the $B$ meson by a pseudoscalar current
$$J_B(x) = \bar b(x) i \gamma_5 q'(x)\,,$$
one considers the correlation function
\beq
  \Pi_\mu^{\rm QCD}(p_B,p'{}) &=& i \int d^4x \,
    e^{- i p_B x}
\,\langle \pi(p'{}) | T[ J_B(x) J_\mu(0)] | 0 \rangle\,.
\eeq
The perturbative calculation of the correlation function requires
large negative values,
$$
p_B^2-m_b^2 \ll 0 \ .
$$
On the other hand, in the sum rule one needs the
imaginary part of the correlation function at values for
$p_B^2$ close to the threshold parameter where
$$
  p_B^2 -m_b^2 = 2 m_b \omega_0
$$
The scaling of the threshold parameter $\omega_0$
has to be chosen as
$$ \Lambda_{\rm QCD} \ll \omega_0 \sim \Lambda \ll m_b \, . $$
such that the ``soft'' scale $\Lambda$ 
can be interpreted as a perturbative scale in the low-energy 
effective theory (SCET$_{\rm II}$).

Let us first consider the diagrams
in Fig.~\ref{plot1}(a,b) which contribute
to the imaginary part of the
correlation function ${\rm Im}[\Pi(\omega_0)]$ at tree-level.
Let us parametrize the momentum of the light quark 
entering  $J_B$ as
\beq
p^\mu_{\bar q}
&=&
\bar u (n_+p') \, \frac{n_-^\mu}{2} - p_\perp^\mu +
      \frac{p_{\bar q}^2- p_\perp^2}{\bar u \, (n_+p')} \frac{n_+^\mu}{2} \,,
\eeq
where $\bar u = 1-u$ denotes the longitudinal momentum fraction,
the transverse momentum components satisfy $n_\pm p_\perp =0$,  and
the virtuality scales as $|p_{\bar q}^2| \sim |p_\perp^2| \sim \Lambda^2$.
Choosing a frame where $(p_\perp p_B)=0$ and $(n_\pm p_B) \sim m_b$,
the denominator of the heavy-quark propagator
in momentum space reads
$$
  (p_B - p_{\bar q})^2 - m_b^2
= 
2 m_b \omega_0
- \bar u (n_-p_B)(n_+p')
 - \frac{p_{\bar q}^2 -p_\perp^2}{\bar u \, (n_+p')} (n_+p_B) + p_{\bar q}^2\,.
$$

Generic parton configurations in the pion are classified as collinear,
and characterized by $u, \bar u = {\cal O}(1)$.
This case corresponds to the diagram in
Fig.~\ref{plot1}(a), and allows us to approximate
\beq
 \mbox{generic} \quad &:&  \qquad (p_B - p_{\bar q})^2 - m_b^2
 = - \bar u \, (n_- p_B)(n_+ p') + {\cal O}(m_b \Lambda) \,,
\eeq
from which we see that:
\begin{itemize}

\item[(i)] the heavy quark propagates on short distances, of order $1/m_b$,
  which would suggest a perturbative treatment
  of the correlation function in terms of $\alpha_s(m_b)$;

\item[(ii)]
  the non-locality of the correlation function is determined by
  $\bar u (n_+ p')$, which only involves the momentum components collinear
  to the light-hadron momentum $p'{}$ which reflects the dominance
  of light-like separations.
\end{itemize}

The situation is different if we approach end-point configurations where
$\bar u $ is small, 
$\bar u = {\cal O}(\Lambda/m_b)$. At tree level this corresponds to
the diagram in Fig.~\ref{plot1}(b), where the spectator quark is soft and,
by momentum conservation, the light quark from the heavy-quark decay is
hard-collinear. Notice that in the heavy-quark limit
this situation is
enforced by the Borel transformation which actually suppresses
$b$\/-quark virtualities $p_b^2-m_b^2 \gg \Lambda m_B$
(in particular it suppresses the configurations in Fig.~\ref{plot1}(a)).
In this case we have
\beq
 \mbox{end-point} \quad &:& \quad
 (p_B - p_{\bar q})^2 - m_b^2 \cr
&& \quad =
2 m_b \omega_0
- \bar u (n_-p_B)(n_+p')
 - \frac{p_{\bar q}^2 -p_\perp^2}{\bar u \, (n_+p')} (n_+p_B)
 + {\cal O}(\Lambda^2)\,.
\eeq
Also,
\begin{itemize}
\item[(i)] the heavy quark propagates at distances
           of order $1/\Lambda$ (as in HQET,
           the virtuality of the
           heavy quark propagator is of order $\Lambda m_b$,
           but the residual heavy-quark
           momentum is of order $\Lambda$).
           These configurations are unsuppressed by the Borel transformation,
           and suggest a perturbative expansion in terms of
           $\alpha_s(\Lambda)$.

\item[(ii)] The correlation function is not necessarily
            dominated by light-like distances, as can be seen by the
            appearance of the (supposed-to-be) sub-leading
            momentum component
            $(n_-p_{\bar q}) = \frac{p_{\bar q}^2-p_\perp^2}
             {\bar u (n_+ p')}$.
            In this case, the convergence of the
            light-cone expansion is not related to
            the $\Lambda/m_b$ power counting. 
            Instead, in the above example,
            one has an expansion based on the formal
            power-counting
            $|p_{\bar q}^2 - p_\perp^2| \sim \Lambda_{\rm QCD}^2
             \ll \Lambda^2$. After Borel transformation it
            translates into an expansion in inverse powers of the
            Borel parameter.
\end{itemize}

Finally, the momentum configurations in Fig.~\ref{plot1}(c,d) contain
the factorizable form-factor contributions. For these situations,
Ball \cite{Ball:2003bf} has shown that the traditional
LCSRs and the QCD factorization approach \cite{Beneke:2000wa}
to the $B \to \pi$ form factors lead to the same result.
Here, the light-cone
separation of the two quark fields in the pion follows from the
structure of the hard-collinear propagators. The perturbative
expansion is thus in terms of $\alpha_s(\mu_{\rm hc})$, and
in this case one observes the suppression of
higher-twist LCDAs of the pion by $1/m_b$.

We conclude that the description of the end-point contributions to
the (``soft'') $B \to \pi$ form factor within LCSR
{\em in the heavy-quark limit}\/
shows similar subtleties as in the
QCD factorization approach \cite{Beneke:2000wa}. This
is related to the fact that in the heavy-quark limit 
the light-cone sum rule is dominated by diagrams like 
in Fig.~\ref{plot1}(b), which do not correspond to
the parton configurations that one would usually
associate with LCDAs of the pion.\footnote{Another
issue is the appearance of large (Sudakov) logarithms
$\ln m_b/\omega_0$ (with $\omega_0 \sim \Lambda$) in (\ref{soft2})
which could be resummed by standard methods a posteriori
\cite{Ball:2003bf}.}
Although, for finite heavy-quark masses, one finds a
numerical stability of the result with respect to the sum rule parameters
and the twist expansion,
the predictions for
the ``soft'' end-point contributions should thus
be interpreted with some care.
Even if one accepts the formal derivation
on the basis of the light-cone expansion
of the correlation function as a sufficiently good approximation,
the problem  can easily be identified on the practical level:
for generic parton configurations (related to the factorizable
contributions in the heavy quark limit) 
the basic non-perturbative object is the moment
\beq
  \langle u^{-1} \rangle_\pi & \equiv &  \int du \, \frac{\phi_\pi(u)}{u}
  = 3 \left( 1 + a_2 + a_4 + \ldots \right)
\eeq
of the leading-twist LCDA of the pion. Here we also indicated the expansion in
terms of Gegenbauer coefficients. Notice that the coefficients in
front of the $a_n$ for this case is 1 for all $n$, and therefore,
in practice, the expansion can be truncated under the assumption that 
the coefficients $a_n$ decrease with $n$.
On the other hand, as explained above, 
the end-point contributions in the heavy-quark limit involve,
for instance, the quantity
\beq
  \phi_\pi'(1) = - 6 \left\{ 1 + 6 a_2 + 15 a_4 + \ldots +
   \mbox{$\left(\begin{array}{c} n+2 \\ 2 \end{array}\right)$} \,  a_n + \ldots \right\} \,.
\label{phipexp}
\eeq
where the coefficients in front of the $a_n$ now grow quadratically
with $n$. For completeness, we also quote the expression for
the value at the symmetric point $u=1/2$, which can be 
constrained from light-cone sum rules \cite{Braun:1988qv},
\beq
 \phi_\pi(1/2) = \frac32 \left(1 - \frac{3}{2} a_2 + \frac{15}{8} a_4 + \ldots
      \right) \, .
\eeq
For this quantity the coefficients in front of the $a_n$
grow as $\sqrt n$ for large $n$.

\begin{figure}[thbp]
\begin{center}
\psfig{file = 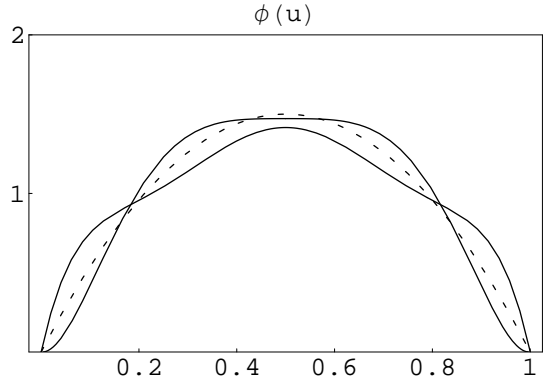}
\end{center}
\caption{The pion LCDA $\phi_\pi(u)$ as a 
function of the longitudinal momentum fraction $u$
for the examples quoted in the text -- solid lines.
For comparison we have also shown the 
asymptotic LCDA $\phi_\pi(u) = 6 u (1-u)$ -- dashed line.}
\label{fig:pidaexp}
\end{figure}

The notation suggests that the two quantities 
$\langle u^{-1} \rangle_\pi $ (or $\phi_\pi(1/2)$) and
$\phi_\pi'(1) $ can be derived from one and the same quantity, $\phi_\pi(u)$,
which could be true if $\phi_\pi(u)$ were {\em exactly}\/ known.
However, in practice, we only have limited information on that function.
Usually, one can constrain the first few Gegenbauer coefficients from
experimental data (for instance the $\pi\gamma$ form factor), or from
sum rules for the first few moments of $\phi_\pi(u)$ 
or for $\phi_\pi(1/2)$ (see \cite{Ball:2005ei}
for a recent discussion and more references).
Let us assume that  we approximate $\phi_\pi(u)$ by keeping only the two terms
$a_2$ and $a_4$ in the Gegenbauer expansion.
As explained above, we can determine the values 
for $a_2$ and $a_4$ only to some
accuracy. The following table compares two not unrealistic cases:
\begin{center}
\begin{tabular}{| c || c c c c | c | c | c |}
\hline
 Case & $\langle u \rangle_\pi$ & $\langle u^2 \rangle_\pi$ &
        $\langle u^3 \rangle_\pi$ &  $\langle u^4 \rangle_\pi$ &
         $\langle u^{-1} \rangle_\pi$ &  $\phi_\pi(1/2)$ & $\phi_\pi'(1)$
\\
 \hline\hline
 $\matrix{ a_2 = -0.05 \\ a_4=-0.05 }$ &
  1/2 & 0.30 & 0.19 & 0.14 &  2.70 & 1.47 & \phantom{-2}0.3
\\
\hline
$\matrix{ a_2 =  \phantom{-} 0.10 \\  a_4 = \phantom{-}0.05 }$ &
  1/2 & 0.31 & 0.21 & 0.16 & 3.45 & 1.42 &  -14.1
\\
\hline
\end{tabular}
\end{center}
The variation of the lowest-order moments 
$\langle u^n \rangle_\pi$ and of $\phi_\pi(1/2)$
in these cases is of the order
of only 10\%, 
whereas neither the sign nor the order of magnitude of $\phi_\pi'(1)$
can be reliably estimated in this way.
In Fig.~\ref{fig:pidaexp} we have illustrated the
shape of the function $\phi_\pi(u)$ for these cases, compared to
the asymptotic LCDA (which corresponds to
$\phi_\pi(u)=6u(1-u)$, $\langle u^{-1}\rangle_\pi=3$, and $\phi_\pi'(1)= -6$).
Notice, that although our choices for 
$a_2$ and $a_4$ are more or less ad hoc, the range
of ``allowed'' or reasonable LCDAs is 
quite similar to the ones obtained from the set
of models proposed in \cite{Ball:2005ei}, 
where the issue of how to constrain $\phi_\pi(u)$
is discussed in more detail.

It follows that for a conservative 
numerical analysis of the LCSRs in the heavy-quark limit, one
should rather consider $\phi'_\pi(1)$ as an independent
non-perturbative parameter. Physically, this is related to the fact that
$\phi_\pi'(1)$ describes completely different momentum configurations than
 $\phi_\pi(u)$ and its moments. Technically, it is impossible to re-construct
the value of $\phi_\pi'(1,\mu_0)$ and its evolution under renormalization from
a finite number of moments of $\phi_\pi(u,\mu_0)$, where $\mu_0$
is a hadronic scale.
This implies that, at the moment,
the predictions for the very end-point contributions in LCSR may have
uncertainties larger than usually considered: in the tree-level
contribution to the form factor $f_+(0) \approx 0.3$, the above
example corresponds
to an uncertainty of about $0.4$, i.e.\ a 100\%
 effect (for $\omega_0 \simeq 1$~GeV).
Notice, that
the situation becomes even worse if we allow for (small)
non-zero values for $a_6$, $a_8$ etc. This implies that
for a quantitative analysis of the theoretical 
uncertainty coming from the hadronic input functions $\phi_\pi$
and $\phi_P$, it is generally not sufficient to only vary $a_2$
and $a_4$, at least in case of the soft form factor
in the heavy-quark limit.
This observation may solve
the apparent problem which has been identified in \cite{Ball:2003bf},
where an unacceptably large value for the 
$B\to\pi$ form factor has been obtained in the
heavy-quark limit.

We should emphasize again that the above criticism applies to the very limit
$m_b \to \infty$ in QCD light-cone sum rules.
For finite heavy-quark masses (which are usually considered)
the sensitivity to the end-point
region appears to be small.
Still, the latter statement requires verification, and
we would like to
propose to study the end-point configurations in QCD light-cone
sum rules more carefully,
and to consider the independent values of $\phi'_\pi(1)$ and $\phi_P(1)$ as
additional sources of systematic uncertainties. It is also true
that for finite heavy-quark masses
it is not mandatory to resum large logarithms $\ln m_b/\Lambda$.
 The prize one has to pay
is that at higher orders in the perturbative expansion the soft form factor
potentially involves
more and more contributions from higher-twist wave functions,
and that the choice of
the renormalization scale has to account for a 
rather large range, say between 1~GeV and $2 m_b$.
Again, as long as the expansion in inverse powers of the Borel parameter
($\Lambda_{\rm QCD}/\Lambda$) is numerically well-behaved and the perturbative
corrections are reasonably small, this does not question the present result
for the central values of the $B \to \pi$
form factor, but may lead to some enhancement of theoretical uncertainties
(see also \cite{Ball:2004ye} for a recent update
 of $B$\/-meson form factors within the sum-rule approach).

On the other hand, in the heavy-quark limit,
a systematic study of radiative corrections does require the
logarithms related to the hard scale $m_b$ to be separated
from those related to  the hard-collinear scale
$\mu_{\rm hc} \sim \sqrt{\Lambda m_b}$.
As we will show below, the effective-theory framework
provides a consistent tool to achieve
the perturbative separation of scales. The sum-rule 
techniques can then be applied to
correlation functions in the effective theory (SCET$_{\rm I}$)
itself. This will automatically lead
to the separation of soft and collinear fields
along  the light-cone, which is a built-in feature of the
SCET Feynman rules and related to the power-counting of
fields and operators in the effective Lagrangian
in terms of a small parameter $\lambda=\sqrt{\Lambda/m_b}$
which formally vanishes in the limit $m_b \to \infty$.
Our formalism does not require an additional twist expansion 
of correlation functions in SCET.
In this framework it is possible to systematically 
control the renormalization-scale dependence of
correlation functions, and to consistently resum large 
logarithms $\ln m_b/\mu_{\rm hc}$ for both,
generic and end-point configurations. The power corrections
from $\Lambda/m_B$\/-suppressed contributions, on the other hand,
are hard to estimate at present. Therefore, the advantages and disadvantages
of the traditional LCSRs and the SCET sum rules, which we are going to 
derive, are to be viewed as complementary.

\section{Sum rules in SCET}

\begin{figure}[tbhp]
\begin{center}
\psfig{file=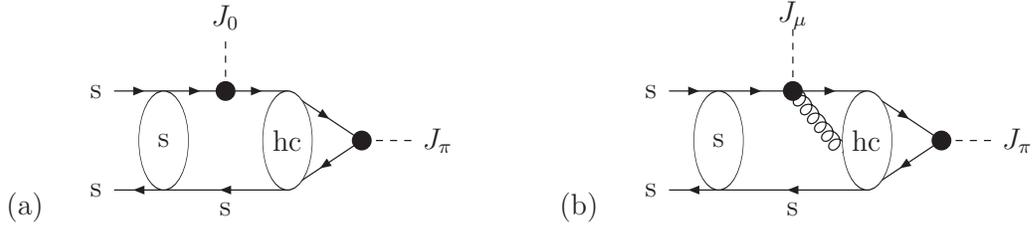, width=0.85\textwidth}
\end{center}
\caption{Contribution to the ``soft'' form factor $\xi_\pi$ (a) and
to the QCD-factorizable contribution from 
hard-collinear spectator scattering (b) to
the SCET sum rule in the heavy-quark limit 
(additional soft spectators are not drawn).
The soft interactions, denoted by the ellipse on
the left-hand side, are parametrized by the light-cone wave function of
the $B$ meson.}
\end{figure}

Within SCET (more precisely SCET$_{\rm I}$, which is the effective theory
describing the interaction of  
soft fields and hard-collinear fields with virtuality $\Lambda m_b$),
the non-factorizable (i.e.\ end-point-sensitive) part
of the $B \to \pi$ form factor in
the heavy-quark limit is described by the current operator
$$
 J_0(0) = \bar \xi_{\rm hc}(0) W_{\rm hc}(0) Y_s^\dagger(0) h_v(0)\,,
$$
where $\xi_{\rm hc}$ is the ``good'' light-cone component of the
light-quark spinor with $\slash n_- \xi_{\rm hc} =0$, 
and $W_{\rm hc}$ and $Y_s^\dagger$
are hard-collinear and soft Wilson lines (the latter appear after one decouples
soft gluons from the leading-power 
hard-collinear Lagrangian, which is convenient for
the following discussion). Finally, $h_v$
is the usual HQET field. As we have seen in the previous Section,
 the heavy quark is nearly on-shell in the end-point
region. In SCET$_{\rm I}$ this is reflected by
the fact that hard sub-processes (virtualities of order $m_b^2$) are already
integrated out and appear in coefficient 
functions multiplying $J_0$. Consequently, the
heavy quark should better be treated 
as an external field and not as a propagating
particle in the correlation function. 
In the SCET sum rules to be derived below,
we therefore will {\em not}\/ introduce 
an interpolating current for the $B$ meson.
Instead, the short-distance
(off-shell) modes in SCET$_{\rm I}$ are the hard-collinear
quark and gluon fields, and therefore 
the sum rules should be derived from a dispersive
analysis of the correlation function
\beq
  \Pi(p') &=& i \int d^4x \,
   e^{i p'{} x} \, \langle 0 | T[ J_\pi(x) J_0(0)] | B(p_B) \rangle\,,
\label{piscet}
\eeq
where $p_B^\mu = m_B v^\mu$, and
\beq
  J_\pi(x) &\equiv& - i \, \bar \psi(x) \, \slash n_+ \gamma_5 \, \psi(x)
\nonumber \\[0.3em]
  & = & - i \, \bar \xi_{\rm hc}(x)  \, 
          \slash n_+ \gamma_5 \, \xi_{\rm hc}(x) - i \,
\left( \bar \xi_{\rm hc} W_{\rm hc}(x)  
       \slash n_+ \gamma_5 Y_s^\dagger q_s(x) + h.c.\right) \,.
\label{eq:Jpiform}
\eeq
Here we denoted quark fields in QCD as $\psi$ and soft and 
hard-collinear quark fields in SCET
as $q_s$ and $\xi_{\rm hc}$, respectively \cite{Beneke:2002ph}.
Notice that soft-collinear interactions require a multi-pole expansion
of soft fields \cite{Beneke:2002ph} which is always understood implicitly. 
We also point out that the effective theory SCET$_{\rm I}$
contains SCET$_{\rm II}$ as its infrared limit (i.e.\ when
the virtuality of the hard-collinear modes is lowered to 
order $\Lambda^2$). For this reason, the hard-collinear fields
which define the interpolating current $J_\pi$ also contain the
collinear configurations which show up as hadronic bound states
(see also the discussion in \cite{Beneke:2003pa}).

According to the discussion in Section~3.4 of the first reference in
\cite{Beneke:2002ph}, the
above expression for $J_\pi$, which has been derived from the
tree-level matching
of QCD onto SCET$_{\rm I}$ will not receive any radiative corrections
since there is no
 hard scale (which could have entered only through
the heavy-$b$\/-quark mass).
The pion-to-vacuum matrix element of the current $J_\pi$ is thus
given as
\beq
  \langle 0| J_\pi | \pi(p')\rangle &=& (n_+p') \, f_\pi \, .
\label{eq:fpidef}
\eeq
In the following we will consider a reference frame where
$p'_\perp=v_\perp=0$ and $n_+v = n_-v = 1$.
In this frame the two independent kinematic variables are
\beq
 (n_+ p'{}) \simeq 2 E_\pi = {\cal O}(m_b) \, , \qquad
 0 > (n_- p'{}) = {\cal O}(\Lambda)\,,
\eeq
with $|n_-p'| \gg m_\pi^2/(n_+p')$.
The dispersive analysis will be performed with respect to
$(n_-p')$ for fixed values of $(n_+p')$.

\subsection{Tree-level result}

\begin{figure}[tbhp]
\begin{center}
\psfig{file=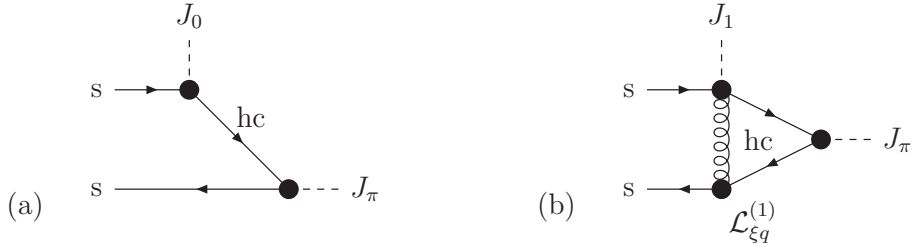, width=0.75\textwidth}
\end{center}
\caption{(a) Leading contribution to the correlation function for the
non-factorizable SCET current $J_0$. (b) Leading contribution for the
factorizable SCET current $J_1$ (see below). }
\label{fig:lead}
\end{figure}

At leading power, the tree-level result, see Fig.~\ref{fig:lead}(a),
for the correlation function involves one hard-collinear quark propagator,
which reads
\beq
  S_F^{\rm hc} = \frac{i}{n_- p'{} - \omega + i\eta}
  \, \frac{\slash n_-}{2}\,,
\eeq
where $\omega = n_- k$, and $k^\mu$ is the momentum of the soft light quark
 that  will end up as the spectator quark in the $B$ meson.
In contrast to the situation discussed in Section~\ref{sec:conv}, 
the light-cone dominance in the SCET sum rules
(now between the soft constituents of the $B$ meson) 
follows from the structure of SCET itself and does 
not constitute an additional assumption.

The perturbative evaluation of the SCET correlation function follows from
contracting hard-collinear fields.
This leads to  matrix elements of operators that are formulated
only in terms of
soft fields, which are separated along the light cone and thus define
light-cone distribution amplitudes for the $B$\/ meson in HQET
\cite{Beneke:2000wa}.
Using the momentum-space representation
of LCDAs for $B$\/ mesons as in \cite{Grozin:1996pq,Beneke:2000wa},
$$
  {\cal M}^B_{\beta\alpha} = - \frac{i f_B m_B}{4}
  \left[ \frac{1+\slash v}{2} \left\{ \phi_+^B(\omega) \slash n_+ +
   \phi_-^B(\omega) \slash n_- + \ldots \right\} \gamma_5
\right]_{\beta\alpha} \,,
$$
we find
\beq
  \Pi(n_- p'{}) & =  &
  f_B m_B \int_0^\infty d\omega \,
   \frac{\phi_-^B(\omega)}{\omega - n_-p'{}-i\eta} \,.
\label{tmp}
\eeq
The considered correlation function in the (unphysical)
Euclidean region thus factorizes into a perturbatively
calculable hard-collinear kernel and a soft light-cone
wave function for the $B$\/-meson, where the convolution variable
$\omega$ represents the light-cone momentum of the spectator quark
in the $B$\/-meson. We will show below that
this factorization still holds after including radiative
corrections in SCET$_{\rm I}$.

The result (\ref{tmp})
already has the form of a dispersion integral in the variable
$n_- p'$
\beq
  \Pi(n_- p')
  &=& \frac{1}{\pi} \int_0^\infty d\omega' \,
      \frac{{\rm Im}[\Pi(\omega')]}{\omega'-n_-p' - i\eta}\,.
\label{disp}
\eeq
The Borel transform with respect to the variable $n_-p'{}$ introduces
the Borel parameter $\omega_M$ and reads
\beq
  \hat B\left[\Pi \right](\omega_M) &= &
      \frac{1}{\pi} \int_0^\infty d\omega' \,
      \frac{1}{\omega_M} \, e^{-\omega'/\omega_M} \,
     {\rm Im}[\Pi(\omega')]
 \\[0.3em]
&=&  f_B m_B \int_0^\infty d\omega \, \frac{1}{\omega_M}
     \, e^{-\omega/\omega_M}  \, \phi_-^B(\omega) \,.
\eeq 
The physical role of the Borel parameter is to enhance the
contribution of the hadronic-resonance region where the virtualities
of internal propagators have to be smaller than the hard-collinear scale. 
The dependence of the sum rule on $\omega_M \sim \Lambda^2/(n_+p')$ will later
be used to quantify the model-dependence of our result. 
On the hadronic side, one can write 
\begin{equation}
\Pi^{\rm HAD}(n_-p')= \Pi(n_- p') \Big|_{\rm res.}+ \Pi(n_- p')
\Big|_{\rm cont.} \,,
\end{equation} 
where the first term represents
the contribution of the pion, while the second takes into account
the role of higher states and continuum above an effective
threshold $\omega_s = {\cal O}(\Lambda^2/n_+p')$. 
Using (\ref{eq:softdef}) and (\ref{eq:fpidef}),
the pion contribution to the dispersive integral is given by 
\beq
  \Pi(n_- p') \Big|_{\rm res.} &= &
   \frac{\langle 0|J_\pi |\pi(p')\rangle \langle \pi(p')|J_0|B(p_B)\rangle}
    {m_\pi^2-p'{}^2}
 \ = \
 \frac{ (n_+p')^2 \, \xi_\pi(n_+p') \, f_\pi}{m_\pi^2-p'{}^2}\,.
\eeq 
Neglecting the pion mass and in the chosen frame where
$p'_\perp=0$, one has \begin{equation} \Pi(n_- p') \Big|_{\rm
res.}=-{(n_+p') \xi_\pi(n_+ p')f_\pi \over n_- p'} \end{equation}
and hence
\begin{equation}
\hat B\left[\Pi \right](\omega_M)\Big|_{\rm res.}={(n_+p')
\xi_\pi(n_+ p')f_\pi \over \omega_M} \,.
\end{equation}

On the other hand, $\Pi(n_- p')
\Big|_{\rm cont.}$ can be written again according to a dispersion
relation analogous to (\ref{disp}) where, assuming global
quark-hadron duality, we identify the spectral density with its
perturbative expression, obtaining: 
\beq
 \hat B\left[\Pi \right](\omega_M) \Big|_{\rm cont.} &= &
      \frac{1}{\pi} \int_{\omega_s}^\infty d\omega' \,
      \frac{1}{\omega_M} \, e^{-\omega'/\omega_M} \,
     {\rm Im}[\Pi(\omega')]
\label{eq:Borel} \\[0.3em]
&=&
  f_B m_B \int_{\omega_s}^\infty d\omega \,
      \frac{1}{\omega_M} \, e^{-\omega/\omega_M} \, \phi_-^B(\omega)\,.
\eeq 
This term can finally be subtracted from (\ref{disp}), giving
the final sum rule 
\beq
  \xi_\pi(n_+p') &= & \frac{1}{f_\pi (n_+p')} \,
   \int_0^{\omega_s} d\omega' \, e^{-\omega'/\omega_M} \, {1 \over
   \pi} {\rm Im}[\Pi(\omega')]
 \,.
\label{sumgeneral} 
\eeq 
At tree level in SCET the previous
equation becomes: 
\beq
  \xi_\pi(n_+p') &= & \frac{f_B m_B}{f_\pi (n_+p')} \,
   \int_0^{\omega_s} d\omega \, e^{-\omega/\omega_M} \, \phi_-^B(\omega)
 \qquad \mbox{(tree level)}\,.
\label{sumexact} 
\eeq
The dependence of $\phi_-^B(\omega)$
occurs on scales $\omega_0 = {\cal O}(\Lambda)$. For small values
of the threshold parameter, $\omega_s \ll \Lambda$, one can
approximate 
\beq
  \xi_\pi(n_+p') &\simeq& \frac{f_B m_B}{f_\pi n_+p'} \, \phi_-^B(0) \,
   \int_0^{\omega_s} d\omega \, e^{-\omega/\omega_M}
\nonumber\\[0.2em]
  &=& \frac{m_B \omega_M}{f_\pi (n_+p')}\left(
1- e^{-\omega_s/\omega_M}\right)  \, f_B \phi_-^B(0)
\, \qquad \mbox{(tree level, approx.)}\,.
\label{sumtree}
\eeq

\subsection{Comments}

\begin{itemize}

\item  The result for $\xi_\pi$ has the correct scaling with
  $\Lambda/m_B$ as obtained from the conventional sum rules or
  from the power counting in SCET.

\item  The factorizable contributions in exclusive
       $B$\/-meson decays usually contain the first inverse moment
       of the $B$\/-meson distribution amplitude $\phi_+^B(\omega)$:
   $$
     \lambda_B^{-1} = \int_0^\infty d\omega \,
         \frac{\phi_+(\omega)}{\omega}
   $$
   On the other hand, the soft form factor is proportional
   to $\phi_-^B(0)$. In the Wandzura-Wilczek (WW) approximation,
   the two quantities are directly related
   \cite{Beneke:2000wa},
   \beq
    \phi_-^B(0) &\simeq &
     \int_0^\infty d\omega \,
      \frac{\phi_+^B(\omega)}{\omega}
    \qquad \mbox{(WW)}\,.
   \label{WW}
   \eeq
   Thus the situation is different from the case discussed
   in Section~\ref{sec:conv},
   where, considering the heavy-quark limit,
   the non-perturbative parameters appearing in the
   hard-scattering and in the soft contributions are not directly
   related to each other. It should, however, be mentioned that
   the above relation is modified by 3-particle Fock states with
   one additional soft gluon in the $B$\/ meson.
   The sensitivity of the soft form factor on
   $\phi_-^B(\omega)$ and three-particle LCDAs has also
   been observed in the context of QCD factorization
   \cite{Hardmeier:2003ig,Lange:2003jz,Beneke:2003pa}.

\item
At tree level, QCD and SCET are indistinguishable,
and therefore our result in (\ref{sumtree}) can
also be derived using the full
QCD Feynman rules \cite{offen}.
For the comparison, one has to identify
$M^2 = \omega_M \, (n_+ p')$ and $s_0 = \omega_s \, (n_+ p')$.
Notice that the organization of radiative
corrections is different in the two frameworks.
Also the behaviour of the form factor as a function of $q^2$
depends on the exact treatment of $(\omega_s,\omega_M)$.
The standard treatment corresponds to $s_0, M^2 = {\rm const.}$,
in which case the soft form factor (at tree-level)
scales as $1/(n_+p')^2$.

One non-trivial issue concerns the hard matching coefficients
between the heavy-to-light currents $\bar q \Gamma_i  Q$ in QCD
and $J_0$ in SCET. They induce a renormalization-scale dependence,
which to NLL
accuracy has the universal evolution  \cite{Bauer:2000yr}:
\beq
  C_i(\mu) &\simeq& 
 C_i(m_b) \, \exp\left[ - \frac{4\pi C_F}{\beta_0^2 \, \alpha_s(m_b)}
  \left(\frac{1}{z}- 1 + \ln z\right) +f_1(z) \right] \, , 
\\[0.2em]
z^{-1} &=& 1+\frac{\beta_0}{2\pi} \, \alpha_s(m_b) \, \ln \frac{\mu}{m_b} \, .
\label{Cevol}
\eeq
The first term in the exponent resums the Sudakov double logarithms
($\alpha_s \, \ln^2 \mu$, \ldots)
between the scales $m_b$ and $\mu$.
The function $f_1(z)$\,,  which takes into account
NLL effects ($\alpha_s \, \ln \mu$,
$\alpha_s^2 \ln^2\mu$, \ldots)\,, can be found
in \cite{Bauer:2000yr}.

\end{itemize}

\subsection{Radiative corrections from hard-collinear loops}

\begin{figure}[hbt]
\begin{center}
 \psfig{file = 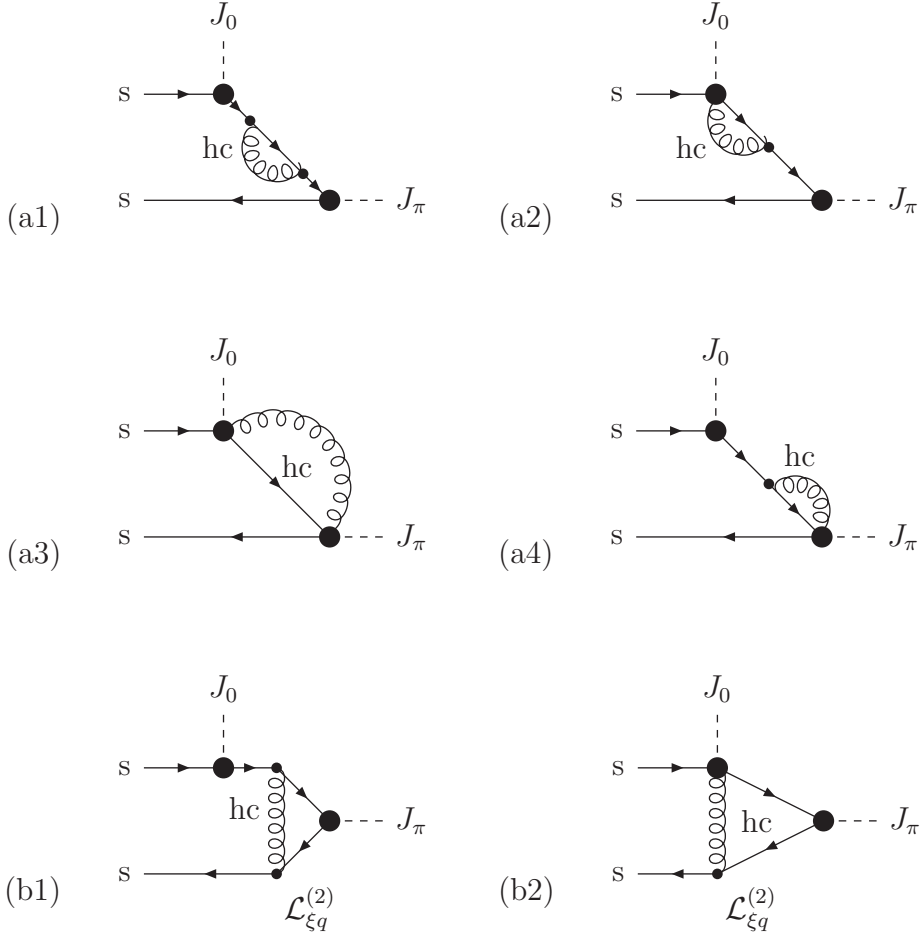}
\end{center}
\caption{Diagrams contributing to the sum rule for $\xi_\pi$
to order $\alpha_s$ with hard-collinear loops and no external
soft gluons.
Diagrams (a2-a4) and (b2) vanish in light-cone gauge $n_+ A_{\rm hc} =0$.
Diagram (a3) vanishes both in light-cone and in Feynman gauge.}
\label{fig:hcloops1}
\end{figure}

In SCET short-distance radiative corrections to 
the correlation function (\ref{piscet})
are represented by hard-collinear loops, 
as shown in Fig.~\ref{fig:hcloops1} for the leading
order in $\alpha_s$.
The diagrams denoted by (a1-a4) and (b1-b2) 
form gauge-invariant subsets, such that before
the Borel transformation the result for the correlation function reads
\beq
   \Pi(n_- p',\epsilon) & =  &
  f_B m_B \int_0^\infty d\omega \, 
  \frac{\phi_-^B(\omega)}{\omega - n_-p'{}-i\eta}
  \left\{ 1 + \frac{\alpha_s C_F}{4\pi} 
  \left( \mbox{(a1-a4)} + \mbox{(b1-b2)} \right) \right\} \,.
\cr &&
\label{hccontri}
\eeq
We obtain
\beq
  \mbox{(a1-a4)} &=&
                  \frac{4}{\epsilon^2} + \frac{1}{\epsilon} \,
   (3 + 4 L_0 - 4 L_1(\omega))
\cr
 && \quad
  + L_0 \left(3 + 2 L_0 - 4 L_1(\omega) \right)
  - L_1 (\omega) \left(3 - 2 L_1(\omega)\right) + 7 - \frac{\pi^2}{3} \,,
\label{a14}
\eeq
and
\beq
  \mbox{(b1-b2)}  &=&
   -\frac{2}{\epsilon^2}
  - \frac{1}{\epsilon}\left(1 + 2 L_0 +
     2 L_1(\omega) \left(1-\frac{n_-p'}{\omega}\right)
  \right)  \cr
 && \
  - L_0 \left(
      1 + L_0 + 2 L_1(\omega) \left(1-\frac{n_-p'}{\omega}\right)
    \right)
\cr
 &&
   - L_1(\omega) (3 - L_1(\omega)) \left(1-\frac{n_-p'}{\omega}\right) 
    - 3 + \frac{\pi^2}{6} \,,
\cr &&
\label{b12}
\eeq
where we have defined the abbreviations
\beq
&&
L_0 =  \ln \left[\frac{\mu^2}{(n_+ p_\pi)(- n_-p' - i\eta)}\right] \, , \qquad
L_1(\omega) =  \ln \left[1 - \frac{\omega}{n_-p'+i\eta}\right] \,.
\eeq
Notice that the hard-scattering diagrams (b1-b2) 
involve a sub-leading term in the SCET Lagrangian,
describing the interactions of soft and hard-collinear 
quarks \cite{Beneke:2002ph}:
\beq
  {\cal L}_{\xi q}^{(2)} &=& \bar q_s W_{\rm hc}^\dagger (in_-D_{\rm hc} +
   i \Slash D_{\perp \rm hc} \, (in_+ D_{\rm hc})^{-1} \,  
   i \Slash D_{\perp \rm hc})
   \frac{\slash n_+}{2} \, \xi + \ldots
\eeq
(The insertion of ${\cal L}_{\xi q}^{(1)}$
vanishes by rotational invariance in the transverse plane.)

\subsection{Renormalization-scale dependence of the correlator}
\label{cancel}

\begin{figure}[tbhp]
\begin{center}
\psfig{file=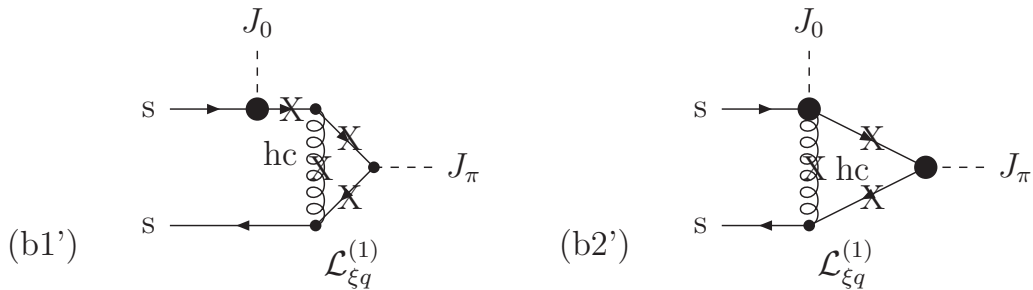, width=0.85\textwidth}
\end{center}
\caption{$\alpha_s$ corrections from hard-collinear loops involving an additional transverse soft gluon.
 The crosses indicate possible insertions of the corresponding interactions with soft gluons from the
sub-leading SCET Lagrangian, ${\cal L}_\xi^{(1)}$
and ${\cal L}_{\rm YM}^{(1)}$.}
\label{Fig:softgluon}
\end{figure}

The renormalization-scale dependence for the correlation
function in (\ref{piscet}) comes from three sources. Let us, for
the moment, focus on the leading double-logarithmic terms:
\begin{itemize}
  \item
    The hard-collinear loop contributions yield
   \beq
    \frac{\alpha_s C_F}{4\pi}  \,
    \frac{\partial}{\partial \ln \mu} \, \mbox{(a1-a4)}^{\overline{\rm MS}} &=&
        \frac{\alpha_s C_F}{4\pi} \left(
         8 \ln \frac{\mu^2}{\mu_{\rm hc}^2} +
         \mbox{$\mu$-independent terms}  \right)\,,
   \eeq
    and
   \beq
    \frac{\alpha_s C_F}{4\pi}  \, 
     \frac{\partial}{\partial  \ln\mu} \, 
    \mbox{(b1-b2)}^{\overline{\rm MS}} &=&
        \frac{\alpha_s C_F}{4\pi} \left(
         -4 \ln \frac{\mu^2}{\mu_{\rm hc}^2} +
         \mbox{$\mu$-independent terms}  \right)\,.
   \eeq

   \item
   The evolution equation of the $B$\/-meson
   LCDA (see \cite{Lange:2003ff,Braun:2003wx}
   and Eq.~(\ref{softdiff}) in Appendix~\ref{app:soft})
   implies
   \beq
     \frac{\partial}{\partial  \ln \mu} \, 
     \int d\omega \, \frac{\phi_-^B(\omega,\mu)}{\omega-n_-p'}
     &=&
     \frac{\alpha_s C_F}{4\pi} \,
       \int d\omega \, \frac{\phi_-^B(\omega,\mu)}{\omega - n_-p'}
      \left(
         - 2 \ln \frac{\mu^2}{\mu_0^2}  +
         \mbox{$\mu$-independent terms}  \right) \,.
   \cr &&
   \eeq
   Here, we have used the fact that
   the anomalous-dimension kernel for $B$-meson LCDAs contains a
   universal term related to the cusp anomalous dimension:
   \beq
     \gamma_\pm(\omega,\omega',\mu) = \Gamma_{\rm cusp}(\alpha_s) \,
       \ln \frac{\mu}{\mu_0} \, \delta(\omega-\omega')  +
         \mbox{$\mu$-independent terms} \,,
   \eeq
    with $\Gamma_{\rm cusp}= 4 + {\cal O}(\alpha_s)$.

   \item The evolution of the $B$\/-meson decay constant $f_B(\mu)$
         in HQET does not contain Sudakov double logs.
\end{itemize}
As required, the resulting scale-dependence of the correlation function
cancels with those of the Wilson coefficients $C_i(\mu)$
in (\ref{Cevol}) to the considered leading logarithmic order
(which involves the Sudakov double logs),
   \beq
     \frac{\partial}{\partial \ln \mu} \, C_i(\mu) &=&
      \frac{\alpha_s C_F}{4\pi} \left( - 2 \ln \frac{\mu^2}{m_b^2}
       + \mbox{$\mu$-independent terms}\right) \, C_i(\mu)\,.
   \label{Cmuapprox}
   \eeq
To show the cancellation of
the sub-leading single-logarithmic terms,
we would have to compute the $\mu$\/-independent
terms in the anomalous-dimension function $\gamma_-(\omega,\omega')$ as
well as the finite $\alpha_s$ corrections from the diagrams shown in
Fig.~\ref{Fig:softgluon}
(the NLL term in $C_i(\mu)$ is known \cite{Bauer:2000yr}).
Notice that to NLL accuracy also the WW approximation
will receive corrections from the
three-particle $B$\/-meson LCDA,  with
one additional soft transverse gluon, see the discussion
in Appendix~\ref{app:soft}.
We will leave the cancellation of single logs and the
corrections from three-particle LCDAs for future investigations.

It is to be stressed that, for the hard-collinear loop diagrams,
the power counting in SCET requires at least one insertion of the
sub-leading Lagrangian ${\cal L}^{(1)}$ for kinematical reasons,
and the coupling of external transverse soft gluons to
hard-collinear particles costs another factor of $\lambda$.
Therefore, in higher orders of perturbation theory, we will not
encounter more than three-particle LCDAs of the $B$ meson. (Of
course, the evaluation of power corrections will require a finite
set of more-particle LCDAs for a given power in $\lambda$. The
role of three-particle Fock states and power corrections to the $B
\to \pi$ form factor is further explored in \cite{offen}. The role
of four-particle Fock states in power corrections to the $B \to
\pi\pi$ amplitudes has already been emphasized in
\cite{Feldmann:2004mg}.)

\subsection{Sum rule at first order in $\alpha_s$}

The NLO sum rule for the soft form factor is obtained
from the dispersion relation~(\ref{disp}), where we now
have to include the ${\cal O}(\alpha_s)$ corrections to the
imaginary part of the correlation function.
In Appendix~\ref{appA}\ we have calculated the imaginary part
of $\Pi(\omega')$ resulting from the hard-collinear loop
contributions in (\ref{a14}) and (\ref{b12}). Here we have also
included the soft contributions related to the hadronic
matrix element which defines the evolution of the
$B$\/-meson LCDA $\phi_-^B(\omega,\mu)$
entering the tree-level sum rule.
Using, for simplicity, the approximation
$\omega_s \ll \Lambda$, we obtain the final sum rule to order $\alpha_s$,
\beq
  \xi_\pi(n_+p',\mu) &\simeq &
  \frac{ f_B(\mu) m_B \, \phi_-^B(0,\mu_0)}{(n_+p')f_\pi}
   \, \int_0^{\omega_s} \,
   d\omega' \, e^{-\omega'/\omega_M}
\nonumber \\[0.3em]
   &&  \Bigg\{
   1 + \frac{\alpha_s C_F}{4\pi}
\Bigg(
   L_0'(2 + L_0')
 + 4
 + \frac{\pi^2}{2}
 + 4 \left(1 + L_0' \right)
     \left(\sigma_B(\mu_0) - \ln \frac{\mu_0}{\omega'} \right)
\nonumber \\[0.3em]
 && \qquad
+ \ln \frac{\mu}{\mu_0} \left(
  - 4 \, \sigma_B(\mu_0)
  - 2 \, \ln\frac{\mu}{\mu_0}
  + 2 \right)
\Bigg) \Bigg\} \,,
\label{sumNLO}
\eeq
where $L_0' = \ln [\frac{\mu^2}{(n_+p') \, \omega'}]$
and $\sigma_B(\mu_0)$ is the logarithmic moment of the
$B$\/-meson LCDA $\phi_+^B(\omega,\mu_0)$ as defined in \cite{Braun:2003wx}.
Notice, that we have neglected the effects from three-particle Fock states
in the $B$\/-meson which would enter the sum rule at order $\alpha_s$, even
in the WW approximation, see the discussion in Appendix~\ref{app:soft}.
In this respect, our NLO prediction is still incomplete.


\subsection{QCD-factorizable contributions from SCET sum rules}

In this paragraph we will show that (at least at leading-order in $\alpha_s$)
our sum-rule approach reproduces the result for the {\em factorizable}\/
form-factor contribution from hard-collinear spectator scattering
\cite{Beneke:2003pa,Beneke:2000wa}.
For this purpose we consider the correlation function
\beq
  \Pi_1(p') &=& i \int d^4x \,
   e^{i p'{} x} \, \langle 0 | T[ J_\pi(x) J_1(0)] | B(p_B) \rangle \,,
\label{piscet1}
\eeq
where
\beq
 J_1 &\equiv &
 \bar \xi_{\rm hc} \,  g \, \slash A^\perp_{\rm hc} \,  h_v
\eeq
determines the factorizable form-factor contribution $\Delta F_\pi$,
see Appendix~\ref{appB}.
The leading contribution is given by the diagram in Fig.~\ref{fig:lead}(b)
which involves the insertion of one interaction vertex from
the order-$\lambda$ soft-collinear Lagrangian
$$
 {\cal L}_{\xi q}^{(1)} 
\ = \ \bar \xi_{\rm hc} \, g_s \Slash A_{\rm hc,\perp} \, q_s + \mbox{h.c.}
$$
The resulting hard-collinear loop integral is UV- and IR-finite,
and the correlator reads
\beq
  \Pi_1(p') &=& - \, g_s^2 \, C_F \,
    \, \int_0^\infty d\omega
    \, {\rm tr}\left[ {\cal M}^B \, \gamma_\mu^\perp \gamma_5 \, \slash n_-
    \gamma^\mu_\perp\right]
  \nonumber \\[0.3em] && {}\times
   \int \frac{d^4l}{(2\pi)^4} \frac{1}{
    [n_-l - \omega + \frac{l_\perp^2}{n_+l} + i \eta]
    [n_-p' + n_-l - \omega + \frac{l_\perp^2}{n_+p'+n_+l}+ i \eta]
[l^2+ i \eta]} \,,
\cr &&
\eeq
where $l^\mu$ is the momentum of the exchanged hard-collinear gluon.
The calculation yields
\beq
  \Pi_1(p') &=& - \frac{\alpha_s C_F}{4\pi}
    (n_+p') \, \int_0^\infty d\omega \,
    \frac{f_B m_B \, \phi_+^B(\omega)}{\omega} \,
    \ln \left[1 - \frac{\omega}{n_-p'+i\eta}\right] \,.
\eeq
In order to perform the Borel transformation and to define
the continuum subtraction we determine the imaginary part
\beq
   \frac{1}{\pi} \, {\rm Im}[\Pi_1(\omega')]
   &=& - \frac{\alpha_s C_F}{4\pi} \,
    (n_+p') \, \int_0^\infty d\omega \,
    \frac{f_B m_B \, \phi_+^B(\omega)}{\omega} \,
    \theta[\omega-\omega'] \,.
\eeq
Inserting this into the dispersion relation analogous to (\ref{disp}),
we obtain
\beq
   \hat B\left[\Pi_1 \right](\omega_M) \Big|_{\rm res.}
 &\simeq& - \frac{\alpha_s C_F}{4\pi} \,
     (n_+p') \left(1 - e^{-\omega_s/\omega_M}\right)
    \int_0^\infty d\omega \,
    \frac{f_B m_B \, \phi_+^B(\omega)}{\omega} \,,
\eeq
where we have neglected a term which is suppressed by
$\omega_s/\Lambda$.
The hadronic expression for the same quantity reads
(see Appendix~\ref{appB}),
\beq
  \hat B\left[\Pi_1 \right](\omega_M) \Big|_{\rm had.}
&=&
  - \frac{\alpha_s C_F}{4\pi} \, \frac{f_\pi m_B^2}{2 \omega_M} \, 
    \Delta F_\pi \,.
\eeq
Comparison with the expression for $\Delta F_\pi$ obtained
in QCD factorization (see \cite{Beneke:2000wa} and (\ref{DeltaF}) in
Appendix~\ref{appB}) implies
\beq
   (n_+ p') \, \omega_M \left(1 - e^{-\omega_s/\omega_M}\right) =
   M^2 \left(1 - e^{-s_0/M^2}\right)
   &\simeq & 4 \pi^2 f_\pi^2 \, , \quad \mbox{and} \quad
   \langle u^{-1} \rangle_\pi \simeq 3 \,.
\label{relation}
\eeq
The first relation is known from the leading-order sum rule for
$f_\pi$ (see for instance \cite{Colangelo:2000dp}).
The second relation states that to the
considered order the pion distribution amplitude can be approximated
by the asymptotic one.\footnote{Notice that $\langle u^{-1} \rangle_\pi$
has been ``calculated'' in fixed-order perturbation theory at
the hard-collinear scale $\mu_{\rm hc} \gg \Lambda$,
$$
 \langle u^{-1} \rangle_\pi = \langle u^{-1} \rangle_{\rm asymptotic}
\,  \left\{ 1 + {\cal O}\left(\alpha_s(\mu_{\rm hc})\right)
\right\} \,.
$$
Higher-orders in perturbation theory would lead to large
logarithms $\ln \Lambda/\mu_{\rm hc}$. Resumming these logarithms
would eventually lead to ``realistic'' distribution amplitudes and
potentially sizeable deviations from the asymptotic
value.}
Actually, in the limit $\omega_s \to \infty$, the result for
the correlation function {\em before}\/ the
integration over $l_\perp$ and $ n_+ l = - \bar u \,  (n_+p')$
can be written as
\beq
   \hat B\left[\Pi_1 \right](\omega_M)  & \propto &
   \frac{\alpha_s C_F}{4\pi}
    \, \left[\int_0^\infty \frac{d\omega}{\omega} \,
    \phi_+^B(\omega) \left(1- e^{-\omega/\omega_M}\right) \right]
    \,
\nonumber \\[0.2em]
 && {} \times
    \left[  \int_0^1 \frac{du}{\bar u} \, \int d|l_\perp|^2 \,  \frac{6}{M^2}
    \,  \exp\left[ -\frac{|l_\perp|^2}{u \bar u \, M^2}\right] \right] \,.
\cr &&
\eeq
One the one hand this explicitly shows the factorization
of the soft and collinear
integrations; on the other hand the structure to the right corresponds
to the asymptotic wave function for the pion
which is often used as a model in phenomenological
applications (with $M^2 = 4 \pi^2 f_\pi^2$ for $s_0 \to \infty$),
see for instance \cite{Lepage:1982gd,Dahm:1995ne,Musatov:1997pu}.

A remarkable feature of the SCET-sum-rule approach to the $B \to
\pi$ form factor is that the ratio of factorizable and
non-factorizable contributions is independent of the $B$\/-meson
wave function to first approximation. From (\ref{sumtree}),
(\ref{WW})  and (\ref{DeltaFres}) in Appendix B we have (at
$q^2=0$) \beq
 \frac{\alpha_s(\mu_{\rm hc}) C_F}{4\pi} \,  \frac{\Delta F(m_B)}{\xi_\pi(m_B)}
   & \simeq &  \frac{\alpha_s(\mu_{\rm hc}) C_F}{2\pi}
   \approx 6 \% \,,
\label{facto}
\eeq
which is in line with the power counting used in QCD factorization
\cite{Beneke:1999br,Beneke:2000wa}, but contradicts the assumptions
of the so-called pQCD approach \cite{Chen:2001pr}
and the results of a recent phenomenological
study in \cite{Bauer:2004tj}.

From these considerations we see that our formalism reproduces the
structure for the QCD-factorizable part of the form factor. The
leading-order analysis suggests a model for the pion light-cone wave
function,
where the pion-distribution amplitude at the hard-collinear
scale is approximated by the asymptotic form, and the
transverse size of the $q\bar q$ Fock state in the pion
is determined by the Borel parameter.
Of course, these approximations should be refined by
including higher-order radiative and non-perturbative
corrections. For the perturbative part the accuracy
can be systematically improved within the SCET framework
\cite{Beneke:2004rc,Hill:2004if}. The non-perturbative uncertainties
are encoded in the pion and $B$\/-meson
distribution amplitudes at the hard-collinear
scale.


\section{Numerical analysis}

In this Section we study the phenomenological implications of
our result concerning the prediction for the $B \to \pi$ form
factor in the heavy-quark limit. Our approach includes several
sources of theoretical uncertainties,
\begin{itemize}
  \item the dependence on the sum-rule parameters $\omega_s$
        and $\omega_M$, including the quality of the
        approximation $\omega_s,\omega_M \ll \Lambda$,
  \item the variation of the renormalization/factorization scale,
  \item the $B$\/-meson distribution amplitude
        $f_B \phi_-^B(\omega,\mu)$,
        in particular its value at $\omega=0$,
\end{itemize}
which we are going to address in turn.

\subsection{Tree-level approximation}

A priori, the threshold parameters $s_0 = \omega_s \, (n_+p')$
and the Borel parameter $M^2 = \omega_M \, (n_+p')$ are
arbitrary. Their order of magnitude can be estimated from
other sum rules, like the one for $f_\pi$
in (\ref{relation}). Together with (\ref{sumtree}) this defines our
leading-order approximation
\beq
  \xi_\pi(n_+p',\mu)
  &\simeq & \frac{4 \pi^2 f_\pi \, f_B(\mu) m_B}{(n_+p')^2} \,
    \phi_-^B(0,\mu)
\, \qquad \mbox{(tree-level, default)} \,.
\label{sumdef}
\eeq
Fixing the input values at the low hadronic (soft) scale,
$\mu_0 = 1$~GeV, we take $\phi_-^B(0,\mu_0) \simeq 2.15$~GeV$^{-1}$
from \cite{Braun:2003wx}, and $f_B(m_b)=180$~MeV (which correspond to
$f_B(\mu_0)=150$~MeV).
With this we obtain, the tree-level estimate
\beq
  \xi_\pi(n_+p',\mu) \approx
  \xi_\pi(n_+p',\mu_0) & \simeq 0.32 \, \frac{m_B^2}{(n_+p')^2} \,.
\eeq
This already has the right order of magnitude that one would
expect from previous studies.

\subsection{Variation of the sum-rule parameters}

It is to be stressed that the leading-contribution
to the SCET sum rule comes from the second term in (\ref{eq:Jpiform})
which corresponds to the asymmetric momentum configuration in the pion.
On the other hand, QCD sum rules for parameters like $f_\pi$
mainly probe the first term in (\ref{eq:Jpiform}),
i.e.\ the symmetric configuration. Therefore, higher-order corrections
to the sum rule for $\xi_\pi$ can be substantially different from
those in $f_\pi$ which should be taken into account in the
estimate of systematic uncertainties.
We may allow for a 30\% violation
of the relation~(\ref{relation}), such that
\beq
  \frac{M^2 \, (1- e^{-s_0/M^2})}{4 \pi^2 f_\pi^2}
  &=& 1.0 \pm 0.3
\eeq
induces a corresponding error in our estimate for $\xi_\pi(n_+p')$.
As a first estimate of the Borel and threshold parameter we start with
$$
  M^2 = 1~{\rm GeV}^2 \,, \qquad
  s_0 = 1~{\rm GeV}^2 \,,
$$
which fulfill the relation~(\ref{relation}).
At $q^2 =0$ this corresponds to
$\omega_M \simeq \omega_s \simeq 0.2$~{\rm GeV}.
These values are reasonably small compared to $\Lambda$.

An additional criterion which may serve as a
self-consistency check of the sum
rule is to compare the (model-dependent) prediction for the
continuum contribution with respect to the full sum rule.
In order not to be too sensitive to the modelling of the continuum
contribution by quark-hadron duality,
one would like to have the ratio of the continuum and
resonance contribution sufficiently small.
To quantify this effect one needs a model for the shape of
the $B$\/-meson distribution
amplitude. For this study we take the simple parameterization suggested in
\cite{Grozin:1996pq}
\beq
  \phi_-^B(\omega) &=& \frac{1}{\omega_0} \, e^{-\omega/\omega_0} \,,
\label{model}
\eeq
with $1/\omega_0 = \phi_-^B(0) = 2.15$~GeV$^{-1}$.
With this we obtain
\beq
  r \equiv \frac{\mbox{continuum}}{\mbox{total}}
       &=&
  \exp\left[ - \omega_s/\Omega_M \right] \,,
\eeq
where for the starting  values
$\Omega_M = \omega_0 \omega_M/(\omega_0 + \omega_M) \approx 0.14$~GeV,
and $r \simeq 0.24$ is reasonably small. 
As mentioned earlier, the scaling for both the Borel parameter 
and for the threshold parameter is fixed
as $\omega_s \sim \omega_M \sim \Lambda^2/M$, and therefore one
always has $r < 1$. We note that before the Borel transformation
(which corresponds to the limit $\omega_M \gg \omega_s$) the
sum rule would in fact be dominated by the continuum contribution,
$r \to 1$.
We see that, as usual, the Borel transformation is crucial to
obtain a reasonable and self-consistent result.

\begin{figure}[tbhp]
  \begin{center}
  \psfig{file=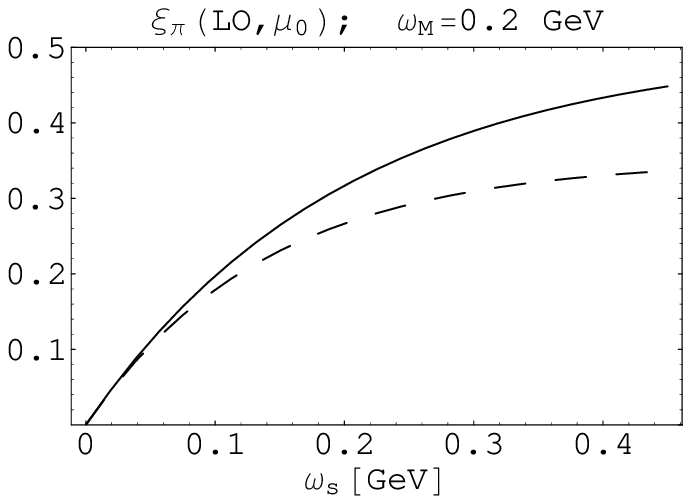, width=0.45\textwidth}
  \psfig{file=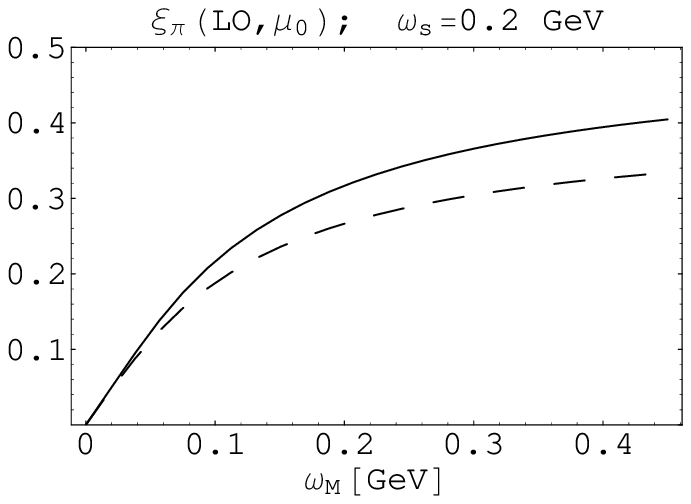, width=0.45\textwidth}
  \end{center}
  \caption{Dependence of the LO prediction for $\xi_\pi(m_B,\mu_0)$
   on the Borel parameters. The solid line refers to the
   approximation $\omega_s,\omega_M \ll \omega_0$; the
   dashed line denotes the exact result, see text.}
  \label{fig:approx}
  \end{figure}

A related question concerns the quality of the approximation
$\omega_s,\omega_M \ll \omega_0 = {\cal O}(\Lambda)$ which
is only marginally fulfilled
for our starting  values of $M^2$ and $s_0$.
In Fig.~\ref{fig:approx} we consider the tree-level sum rule
and compare the ratio of the approximate result (\ref{sumtree})
and the exact formula (\ref{sumexact})
as a function of $\omega_s$ and $\omega_M$.
Here we employed again the model (\ref{model}).
One observes that the approximate formula (solid line) tends to
overestimate the exact result (dashed line) by about 15\% at
the ``default'' values for the Borel parameters. The plots also show
the dependence of $\xi_\pi(m_B,\mu_0)$ on the sum rule parameters
$\omega_s$ and $\omega_M$ which is substantial.

\begin{figure}[tbhp]
  \begin{center}
  \psfig{file=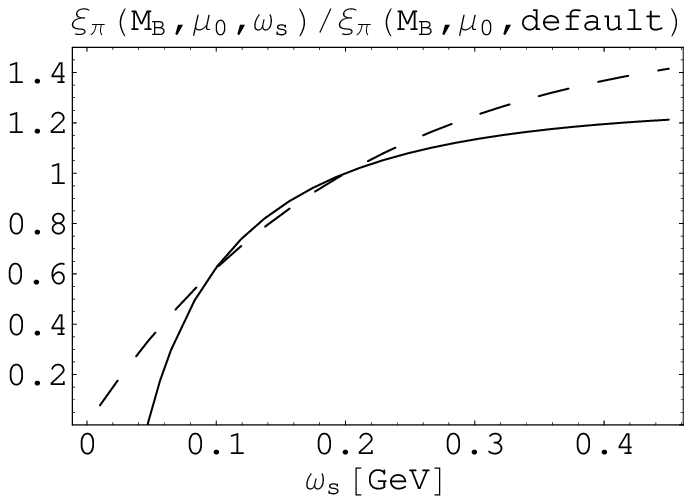, width=0.45\textwidth}
  \psfig{file=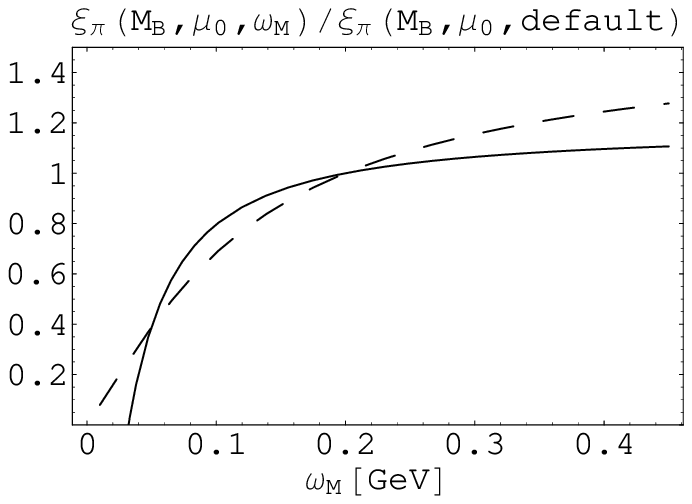, width=0.45\textwidth}
  \end{center}
  \caption{Dependence of the form-factor prediction on the
   sum rule parameters at LO (dashed line) and NLO (solid line),
   normalized to the ``default'' value with $\omega_s = \omega_M = 0.2$~GeV.}
  \label{fig:borelNLO}
  \end{figure}

At the NLO level
the stability  of the sum rule with respect
to variation of the sum-rule parameters is improved
in particular for $(\omega_s,\omega_M) > 0.2$~GeV,
see Fig.~\ref{fig:borelNLO}.
From these considerations we deduce an uncertainty of $(+20\%,-30\%)$
from the variation of $\omega_s$ and $\omega_M$ (a complete 
stability analysis is reserved for further studies 
when all contributions are known and
the approximation on $\omega_{s,M}$  is lifted).

\subsection{Renormalization-scale dependence}

At tree level the question of renormalization-scale dependence
reflects itself in the ambiguous choice of reference scale for the
hard Wilson coefficients, and the $B$-meson distribution
amplitude. Notice that only the product $C_i(\mu)\,
\xi_\pi(n_+p',\mu)$ should be renormalization-scale invariant. A
``reasonable'' choice of scale could be of the order of the
hard-collinear scale, $\mu_{\rm hc} \simeq 1.5$~GeV or of the soft
scale $\mu_0 \simeq 1$~GeV. Below, we will therefore vary the
scale parameter in a sufficiently large range, $\mu_0/2 \leq \mu
\leq 2 \mu_{\rm hc}$.

\begin{figure}[tbhp]
  \begin{center}
  (a) \psfig{file=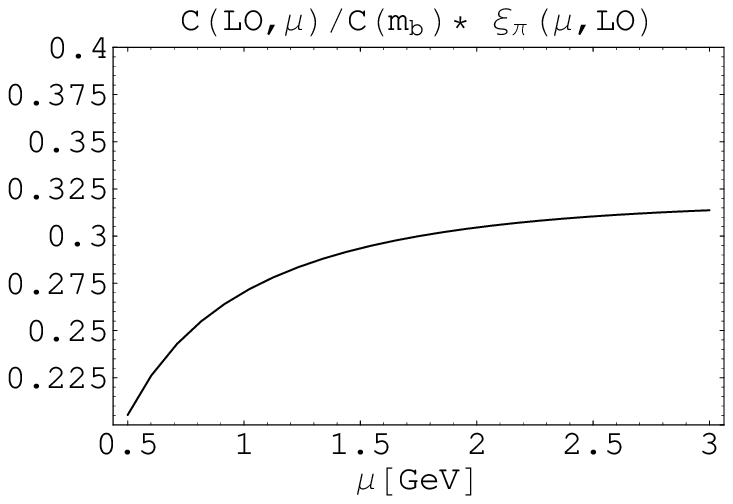, width=0.4\textwidth}
  \hspace{2em}
  (b) \psfig{file=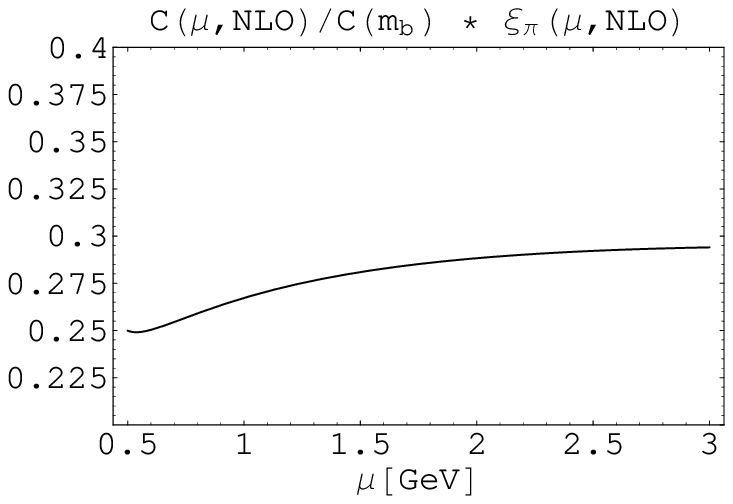, width=0.4\textwidth}
  \end{center}
  \caption{Study of the renormalization-scale dependence of the
    product $C_i(\mu)/C_i(m_b) \, \xi_\pi(m_B,\mu)$
    (a) at tree-level and using LL for the Wilson coefficients,
    (b) at NLO and using NLL for the Wilson coefficients.}
  \label{fig:mudep}
  \end{figure}

In the tree-level approximation the renormalization-scale
dependence of the product $C_i(\mu) \, \xi_\pi(n_+p',\mu)$ is
solely coming from the Wilson coefficients. For the range of
scales indicated above we obtain 
\beq
  \frac{C_i(\mu)}{C_i(m_b)} \big|_{LL}
 &=& 0.85^{+0.14}_{-0.20} \,,
\eeq 
where the central value corresponds to $\mu=\mu_0$. The
suppression from the resummation of the Sudakov logarithms in
$C_i(\mu)$ is moderate, of the order of at most 15-35\%, but the
related scale uncertainty is sizeable. This is also shown in
Fig.~\ref{fig:mudep}(a).

At NLO, we have to take into account the corrections from the
hard-collinear loop diagrams, as well as the soft evolution
effects from the $B$\/-meson distribution amplitude. We also take
into account the NLL approximation for the Wilson coefficients.
Using (\ref{sumNLO}) with $\sigma_B(\mu_0) \simeq 1.4$ taken from
\cite{Braun:2003wx}, we obtain 
\beq
    \frac{C_i(\mu)}{C_i(m_b)} \, \xi_\pi(m_B,\mu) \Big|_{NLO}
 &=& 0.27^{+0.02}_{-0.02} \,,
\eeq 
where the error denotes the renormalization-scale uncertainty
only. Here, the central value corresponds to $\mu_0=1$~GeV, and we
have used $\alpha_s(m_b) \simeq 0.22$ for $m_b = 4.6$~GeV (we use
2-loop running for $\alpha_s$, but for simplicity we have kept
$n_f = 4$ active quark flavours over the whole range of $\mu$).
The situation is also illustrated in Fig.~\ref{fig:mudep}(b).
Notice that the scale-dependence is significantly reduced,
although we have to repeat that our NLO result is incomplete
because of the missing contributions from 3-particle Fock states.

\subsection{The value of $f_B \phi_-^B(0)$ and the final estimate}

For the product of $f_B$ and $\phi_-^B(0)$ we will use
the input values $f_B(m_b) = 180\pm 30$~MeV and
$\phi_-^B(0,\mu_0) = 2.15 \pm 0.5  $~GeV$^{-1}$.
This results in another 25\% uncertainty.
Combining everything we obtain our final NLO prediction
\beq
\frac{C_i(\mu)}{C_i(m_b)} \cdot  \xi_\pi(m_B,\mu)
&=&
  0.27 \pm 0.02 \big|_\mu \
             \pm 0.07 \big|_{f_B\phi_-} \
             {}^{+0.05}_{-0.08} \big|_{\omega_{M,s}} \
             {}^{+0.00}_{-0.04} \big|_{\rm approx.} \,,
\eeq
which compares well with other estimates for the
$B \to \pi$ form factor in {\em full}\/ QCD
(see for instance the recent sum rule result \cite{Ball:2004ye}).
Again, this is consistent with the fact that the factorizable corrections in
(\ref{facto}) are small.
The total parametric uncertainties are still large.
In particular,
the $B$\/-meson LCDA represents an almost irreducible uncertainty,
at the moment.\footnote{This lead the authors of \cite{offen}
to interpret these kind of sum rules as an independent estimate of
$\phi_-^B(0) \simeq \lambda_B^{-1}$ for a given input value
 of  the $B \to \pi$ form factor.}
However, our knowledge about the $B$\/-meson distribution amplitudes
and the appropriate choice of sum-rule parameters should improve
in the future, when we apply our formalism
to various exclusive $B$ decay amplitudes and compare with
experimental data.


\section{Summary and outlook}

We have shown how to derive light-cone sum rules for exclusive
$B$\/-decay amplitudes at large recoil
within the framework of the soft-collinear
effective theory (SCET). Our formalism defines a consistent
scheme to calculate both  factorizable and non-factorizable
contributions to exclusive $B$ decays as a power expansion
in $\Lambda/m_b$. The non-perturbative information is encoded
in terms of light-cone wave functions of the $B$ meson, and
sum-rule parameters related to the interpolating currents for
the light-hadron system in the final state.
Here, our approach resembles the treatment of inclusive $B$ decays
in the so-called shape-function region \cite{inclusive}, where
one factorizes the forward-scattering amplitude into perturbatively
calculable hard coefficient functions, hard-collinear jet functions
in SCET, and soft shape functions of the $B$ meson. The exclusive
sum rule starts from a correlation function between the $B$ meson
and the vacuum, which then factorizes in a quite similar way.
(In this sense our formalism also contains the exclusive leptonic
radiative decays \cite{radiative} as a special case.)

As an explicit example we have studied the factorizable and
non-factorizable contributions to the $B \to \pi$ form factor at
leading power in $\Lambda/m_b$. For the factorizable part of the
form factor our SCET sum rule, the conventional light-cone sum
rule \cite{Ball:2003bf} and the QCD factorization approach \cite{Beneke:2000wa}
coincide. Our central value for the ``soft''/non-factorizable
$B \to \pi$ form factor is consistent with other estimates for
the $B \to \pi$ form factor in full QCD. In particular,
we find that to first approximation, the {\em ratio}\/ of factorizable
and non-factorizable contributions is independent of the $B$\/-meson
wave function and small (formally
of order $\alpha_s$ at the hard-collinear scale, numerically of
the order of 5-10\%).
We have also seen that the suppression of the soft form factor from
Sudakov effects is moderate. We thus confirm the power-counting adopted
in the QCD-factorization approach. Furthermore, we have provided
arguments for why the heavy-quark limit of the traditional
light-cone sum rules fails to reproduce the phenomenological value
of the soft form factor \cite{Ball}.
In this context our observations may also trigger a more sophisticated
discussion of theoretical uncertainties from light-cone sum rules at
{\em finite}\/ heavy-quark mass.

The improvement of the SCET sum rule for the particular case of
the $B \to \pi$ form factor and the extension to other (more
complicated, and perhaps more interesting) decays requires
a better understanding of both, the size and the
renormalization-group behaviour,
of the light-cone wave functions for
higher Fock states in the $B$ meson.
Even to leading power, the soft $B \to \pi$ form
factor receives contributions from a three-particle Fock state,
which has been neglected in this work. Power corrections from
long-distance annihilation or penguin topologies to charmless $B$ decays
involve a 4-quark Fock state in the $B$ meson, etc.
In the future, we  hope to improve our knowledge on
these issues by confronting the SCET-sum-rule
approach to experimental data for various exclusive $B$ decays.

\section*{Acknowledgements}

T.F.\ wishes  to thank the people at the INFN in Bari for the kind hospitality
during his stay, when this work was initiated.
We are grateful to Volodja Braun, Pietro Colangelo and
Alex Khodjamirian for helpful comments and constructive criticism. 
We further thank the authors of
\cite{offen} for discussing their results prior to publication.

\clearpage

\appendix

\section{Calculation of the imaginary part of $\Pi(p')$}

\label{appA}

The $\alpha_s$ corrections to the correlation function
have the general structure
\beq
  \Pi(\omega',\epsilon) &=& {\cal N} \int_0^\infty d\omega \,
      \frac{f(\omega,\omega',\epsilon)}{\omega-\omega'-i\eta}
      \, \phi_-^B(\omega) \,,
\eeq 
where ${\cal N}=f_B m_B \displaystyle{\alpha_s C_F \over 4
\pi}$ and we have indicated that the kernel $f(\omega,\omega')$ is
dimensionally regularized. To separate the UV-region of that
integral we introduce an auxiliary scale $\mu_0 \sim \Lambda \gg
\omega_s \geq \omega'$ and write 
\beq
  \Pi(\omega',\epsilon) &\equiv& \Pi_<(\omega',\mu_0,\epsilon)
    + \Pi_>(\omega',\mu_0,\epsilon) \,,\label{decomp}
 \\[0.4em]
\Pi_<(\omega',\mu_0,\epsilon) &=&
 {\cal N}  \int_0^{\mu_0}
      d\omega \,
      \frac{ f(\omega,\omega',\epsilon)}{\omega-\omega'-i\eta} \,
      \phi_-^B(\omega)
\,, \\[0.3em]
\Pi_>(\omega',\mu_0,\epsilon) &=&
 {\cal N}\int_{\mu_0}^\infty d\omega \,
      \frac{ f(\omega,\omega',\epsilon)}{\omega-\omega'} \,
      \phi_-^B(\omega) \,.
\eeq 
For the second term in (\ref{decomp}) the imaginary part is
given by 
\beq {\rm Im}\left[ \Pi_>(\omega',\mu_0,\epsilon) \right]
&=& {\cal N} \int_{\mu_0}^\infty d\omega \,
      \frac{ {\rm Im}[f(\omega,\omega',\epsilon)]}{\omega-\omega'} \,
      \phi_-^B(\omega) \,.
\eeq
To treat the would-be
singular contribution from $\omega = \omega'$ in the first term in
(\ref{decomp})
we write as usual
\beq
  \Pi_<(\omega',\mu_0,\epsilon) &=& {\cal N}\int_0^{\mu_0} d\omega \,
      \frac{f(\omega,\omega',\epsilon)}{\omega-\omega'-i\eta}
       \, [\phi_-^B(\omega)-\phi_-^B(\omega')]
   +  \phi_-^B(\omega') \,{\cal N} \int_0^{\mu_0} d\omega \,
      \frac{f(\omega,\omega',\epsilon)}{\omega-\omega'-i\eta}
\nonumber \\[0.5em]
 &\equiv& {\cal N} \int_0^{\mu_0} d\omega \,
      \left[\frac{f(\omega,\omega',\epsilon)}{\omega-\omega'}
      \right]_+  \phi_-^B(\omega)
   \ +  \  \phi_-^B(\omega') \,{\cal N} \int_0^{\mu_0} d\omega \,
      \frac{f(\omega,\omega',\epsilon)}{\omega-\omega'-i\eta} \,.
\label{decomp2}
\eeq
From this the imaginary part may be calculated via
\beq
  {\rm Im}[\Pi_<(\omega',\mu_0,\epsilon)] &=&
     {\cal N} \int_0^{\mu_0} d\omega \,
      \left[\frac{ {\rm Im}[f(\omega,\omega',\epsilon)]}{\omega-\omega'}
      \right]_+  \phi_-^B(\omega)
\nonumber \\[0.3em]
  && \ +  \  \phi_-^B(\omega') {\cal N} \left\{
     \pi \, {\rm Re}[f(\omega',\omega',\epsilon)]
   +  {\cal P} \int_0^{\mu_0} d\omega \,
      \frac{ {\rm Im}[f(\omega,\omega',\epsilon)]}{\omega-\omega'}
      \right\} \,,
\eeq
where ${\cal P}$ denotes the principal-value description. Notice that
in general the integral has to be calculated {\em before}\/ the
$\epsilon$\/-expansion.

\subsection{Diagrams a1-a4}

In this case the singular contribution from $\omega =\omega'$ is
dimensionally regularized. Thus, the second term in
(\ref{decomp2}) can be directly calculated, while the first term
in  (\ref{decomp2}) just vanishes in the region $ 0 \leq \omega
\leq \omega'$, if for $\omega \leq \omega' \ll \Lambda$, one
approximates $\phi_-^B(\omega) \simeq \phi_-^B(\omega')
  \simeq \phi_-^B(0)$ -- as in the tree-level result.
Then we get 
\beq
 \frac{1}{\pi}
 \, {\rm Im}[\Pi_<^{(a)}(\omega',\mu_0)]
&\simeq & {\cal N} \left[
  \frac{4}{\epsilon^2}
 + \frac{3 + 4 L_0'(\omega')}{\epsilon}
 + L_0'(\omega') (3 + 2 L_0'(\omega')) + 7 - \pi^2 \right] \, \phi_-^B(0) \,,
\cr &&
\eeq 
where $L_0'(\omega')=\mbox{ln}(\mu^2/(n_+p'\,\omega'))$.
${\rm Im}[\Pi_>^{(a)}(\omega',\mu_0)]$ vanishes and so we have
after $\overline{\rm MS}$\/-subtraction 
\beq
  \frac{1}{\pi}
 \, {\rm Im}[\Pi^{(a)}(\omega')] \simeq & {\cal N}
  \left[ L_0'(\omega') (3 + 2 L_0'(\omega')) + 7 - \pi^2 \right] \phi_-^B(0) 
\,.
\label{ares} 
\eeq

\subsection{Diagrams b1+b2}

We obtain (normalized
to the tree-level result
in units of $\alpha_s C_F/4\pi$)
\beq
 {\rm Re}\left[
   f^{(b)}(\omega',\omega',\epsilon)\right]
&=&
 -\frac{2}{\epsilon^2}
 - \frac{1+ 2 L_0'(\omega')}{\epsilon}
 - L_0'(\omega')(1 + L_0'(\omega')) - 3 + \frac{7\pi^2}{6}
\eeq as well as \beq
 \frac{1}{\pi} \, {\rm Im}\left[
   f^{(b)}(\omega,\omega',\epsilon)\right]
&=&
\left( - \frac{2}{\epsilon}
  -  1 - 2 L_0'(\omega') + 2 \, \frac{\omega'-\omega}{\omega} \,
       \ln \left[1 -  \frac{\omega}{\omega'}\right]
 \right) \theta[\omega'-\omega]
\nonumber\\[0.2em]
&& {} +  \frac{\omega'}{\omega}
 \left(
 - \frac{2}{\epsilon}
  -  1 - 2 L_0'(\omega')
  + 4 \, \frac{\omega'-\omega}{\omega'}
  \left( \frac{1}{\epsilon} + 1 + L_0'(\omega') \right)
 \right) \theta[\omega-\omega'] \,,
\cr &&
\eeq
and in particular
\beq
 \frac{1}{\pi} \, {\rm Im}\left[
   f^{(b)}(\omega,\omega' \to 0,\epsilon)\right]
&=&
  - 4
  \left( \frac{1}{\epsilon} + 1 + L_0'(\omega') \right) \,.
\eeq 
The principal-value integral is given by (for $\mu_0 \gg
\omega'$) 
\beq
  \frac{1}{\pi} \,  {\cal P} \int_0^{\mu_0} d\omega \,
      \frac{ {\rm Im}[f^{(b)}
(\omega,\omega',\epsilon)]}{\omega-\omega'}
&\simeq & -\frac{4}{\epsilon}
  \ln \frac{\mu_0}{\omega'}
  - 4 \, (1+L_0'(\omega')) \, \ln \frac{\mu_0}{\omega'}
  + \frac{\pi^2}{3} \,.
\eeq
After $\overline{\rm MS}$ subtraction and for
$\omega \sim \mu_0 \gg \omega'$ we can approximate
\beq
 \frac{1}{\pi} \, {\rm Im}\left[\Pi_>^{(b)}(\omega',\mu_0) \right]
 & \simeq &
 - 4 \, (1 + L_0'(\omega')) {\cal N} \int_{\mu_0}^\infty
   \frac{d\omega}{\omega} \, \phi_-^B(\omega) \,,
\eeq
and
\beq
  \frac{1}{\pi} \,
 {\rm Im}\left[\Pi_<^{(b)}(\omega',\mu_0) \right]
 &\simeq &
  - 4 \, (1 + L_0'(\omega')) {\cal N} \int_0^{\mu_0}
   \frac{d\omega}{[\omega]_+} \, \phi_-^B(\omega)
\cr + \phi_-^B(0) &{\cal N}& \left(  - L_0'(\omega')(1 +
L_0'(\omega'))
 - 4 \, (1+L_0'(\omega')) \, \ln \frac{\mu_0}{\omega'}
  - 3 + \frac{3\pi^2}{2} \right) \,.
\eeq
The two contributions can be combined, using integration
by parts, which results in the final result
\beq
   \frac{1}{\pi} \,
 {\rm Im}\left[\Pi^{(b)}(\omega') \right]
 &\simeq &
  4 \, (1 + L_0'(\omega')) {\cal N}\int_0^\infty
   d\omega \, \ln \left[\frac{\omega}{\mu_0} \right]
   \left[\frac{d}{d\omega} \, \phi_-^B(\omega) \right]
\cr  + \phi_-^B(0) &{\cal N}& \left\{  - L_0'(\omega')(1 +
L_0'(\omega'))
 - 4 \, (1+L_0'(\omega')) \, \ln \frac{\mu_0}{\omega'}
  - 3 + \frac{3\pi^2}{2} \right\} \,.
\eeq
Notice that in the WW approximation the first term
can be re-written using
$\phi_-^B{}'(\omega) \simeq -\phi_+^B(\omega)/\omega$.
The integral then defines the logarithmic moment of
$\phi_+^B(\omega)$ which enters the evolution equation
for the $B$\/-meson LCDAs.
\beq
   \frac{1}{\pi} \,
 {\rm Im}\left[\Pi^{(b)}(\omega') \right]
 \simeq  \phi_-^B(0){\cal N} &\Bigg\{
  &4 \, (1 + L_0'(\omega')) \left( \sigma_B(\mu_0) -\ln \frac{\mu_0}{\omega'} \right)
 \nonumber \\ &-& L_0'(\omega')(1 + L_0'(\omega'))
  - 3 + \frac{3\pi^2}{2} \Bigg\} \,.
\label{bres}
\eeq
The corresponding hadronic parameter $\sigma_B(\mu_0)$
has been estimated in \cite{Braun:2003wx}.

\subsection{Radiative corrections to soft matrix element}

\label{app:soft}

In this Section we discuss the
radiative corrections to the soft matrix element that
defines the $B$\/-meson distribution amplitude $\phi_-^B(\omega,\mu)$.
As already mentioned, in this work we are sticking to the
Wandzura-Wilzcek approximation (\ref{WW}). However, because of
the non-multiplicative nature of the evolution equations
for $B$\/-meson distribution amplitudes \cite{Lange:2003ff},
the WW approximation is not stable under evolution, in
other words, it can only be valid at a particular scale,
say at a low (soft) scale $\mu_0 \sim \Lambda$
\beq
 && \phi_-^B(0,\mu_0) \simeq \lambda_B^{-1}(\mu_0) \equiv \int_0^\infty d\omega
    \, \frac{\phi_-^B(\omega,\mu_0)}{\omega}
\nonumber \\[0.2em]
  &\Rightarrow&
  \phi_-^B(0,\mu) \simeq \lambda_B^{-1}(\mu)
     \left(1 +
    \frac{\alpha_s C_F}{4\pi} \,
    \ln \frac{\mu}{\mu_0} \ \Delta^B_{\rm WW}(\mu_0) + {\cal O}(\alpha_s^2)
    \right) \,,
\eeq
where $\Delta^B_{\rm WW}(\mu_0)$ parametrizes the unknown
correction term, coming from the evolution of the
three-particle distribution amplitude.

Apart from this missing piece, we can determine the
soft evolution of $\phi_-^B(\omega',\mu)$ from the corresponding
equation for $\phi_+^B(0,\mu)$ in \cite{Lange:2003ff,Braun:2003wx}
Let us, for simplicity, work in fixed-order perturbation theory.
We then have
\beq
 && \phi_-^B(0,\mu) - \phi_-^B(0,\mu_0)
\nonumber \\[0.2em]
  &\simeq &
   \phi_-^B(0,\mu_0)
   \, \frac{\alpha_s C_F}{4\pi} \, \ln \frac{\mu}{\mu_0}
   \left\{ - 4 \, \sigma_B(\mu_0) -2 \ln\frac{\mu}{\mu_0} + 2
   + \Delta^B_{\rm WW}(\mu_0) \right\}
   + {\cal O}(\alpha_s^2) \,.
\cr &&
\label{softdiff}
\eeq
Here, compared to \cite{Braun:2003wx} we kept
the double-logarithmic term $\ln^2 [\mu/\mu_0]$.

\subsection{Imaginary part of the correlator at NLO}

Combining the terms in (\ref{ares},\ref{bres},\ref{softdiff}),
one obtains the final result for the imaginary part of
the correlator at NLO (for $\phi_-^B(\omega',\mu_0) \simeq
\phi_-^B(0,\mu_0)$)
\beq
  \frac{1}{\pi} \,
  {\rm Im}\left[\Pi^{(NLO)}(\omega',\mu) \right]
 &\simeq&  f_B(\mu) \, m_B \, \phi_-^B(0,\mu_0) \Bigg\{
   1 + \frac{\alpha_s C_F}{4\pi}
   \Bigg(
\nonumber \\[0.2em]
&&
 + L_0'(\omega')(2 + L_0'(\omega'))
 + 4
 + \frac{\pi^2}{2}
 + 4 \left(1 + L_0'(\omega') \right)
     \left(\sigma_B(\mu_0) - \ln \frac{\mu_0}{\omega'} \right)
\cr && \quad
+ \ln \frac{\mu}{\mu_0} \left(
  - 4 \, \sigma_B(\mu_0)
  - 2 \, \ln\frac{\mu}{\mu_0}
  + 2
  + \Delta^B_{\rm WW}(\mu_0) \right)
\Bigg) \Bigg\}
\nonumber \\[0.2em]
 && {} + \frac{\alpha_s C_F}{4\pi} \,
    \Delta_{b\bar q g}^B(\mu,\mu_0,\omega') \,,
\eeq 
where the last term $\Delta_{b\bar q g}^B$ denotes the
missing contributions from the three-particle LCDA of the
$B$\/-meson which enter through the diagrams in
Fig.~\ref{Fig:softgluon}. In the above formula, the dependence on
the soft scale $\mu_0$ cancels by means of the (assumed) evolution
equation for $\phi_-^B(0,\mu)$. The dependence of ${\rm
Im}\left[\Pi(\omega',\mu) \right]$ on the factorization scale
$\mu$ should be the same as for the soft form factor
$\xi_\pi(n_+p',\mu)$, and has to cancel with the scale-dependence
of the Wilson coefficients in (\ref{Cevol}). For the double
(Sudakov) logarithms we show this explicitly in
Section~\ref{cancel}. To prove this for the single logarithms we
would need to determine the still unknown functions $\Delta_{\rm
WW}$ and $\Delta_{b\bar q g}^B$ which is left for future studies.
In this case, the evolution equation for our explicit expression
$\xi_\pi(n_+p',\mu)$ can be used all the way down to the scale
$\mu_0$, and the intermediate hard-collinear scale does not appear
explicitly anymore. This is in  agreement with the general
conclusions in \cite{Lange:2003pk}.

\section{The factorizable form-factor contribution $\Delta F_\pi$}

\label{appB}

Let us, for simplicity,  consider the tree-level
matching for heavy-to-light currents in SCET$_{\rm I}$,
using light-cone gauge ($W_{\rm hc} = Y_{\rm s} =1$).
From Eqs.~(124,125) in \cite{Beneke:2002ph} we have
\beq
  \bar \psi \, \Gamma_i \, Q & \longrightarrow &
   \bar \xi_{\rm hc} \, \Gamma_i \, h_v \cr
 && - \frac{1}{n_+ p'} \, \bar \xi_{\rm hc} \, g \Slash A^\perp_{\rm hc} \,
      \frac{\slash n_+}{2} \, \Gamma_i \, h_v
    - \frac{1}{m_b} \, \bar \xi_{\rm hc} \, \Gamma_i \,
      \frac{\slash n_-}{2} \, g \Slash A^\perp_{\rm hc} \, h_v + \ldots
\eeq
where we have neglected terms that give rise to sub-leading
form-factor contributions. The spectator-scattering terms can be
identified by comparing $B \to \pi$ form factors for different
Dirac structures. Let us first consider the scalar current, $\Gamma_i =1$,
for which the tree-level matching reads
\beq
  \bar \psi \, Q & \longrightarrow &
   \bar \xi_{\rm hc} \, h_v
   - \frac{1}{n_+ p'} \, \bar \xi_{\rm hc} \, g \Slash A^\perp_{\rm hc} \,
      \frac{\slash n_+}{2} \, h_v
 \ = \  J_0 - \frac{1}{n_+ p} \, J_1 \,,
\eeq
where we have defined the factorizable current that
appears in Fig.~\ref{fig:lead}(b),
\beq
 J_1 &\equiv &
 \bar \xi_{\rm hc} \,  g \, \slash A^\perp_{\rm hc} \,  h_v \,,
\eeq
and used that $ \slash n_+ = 2 \slash v - \slash n_-$ and
$\bar \xi_{\rm hc} \slash n_-=0$. Similarly, for a vector
current projected with $n_+^\mu$, we obtain
\beq
  \bar \psi \, \slash n_+ \, Q & \longrightarrow &
   \bar \xi_{\rm hc} \, \slash n_+ \, h_v
    - \frac{1}{m_b} \, \bar \xi_{\rm hc} \,
      \frac{\slash n_+ \slash n_-}{2}
  \, g \Slash A^\perp_{\rm hc} \, h_v + \ldots
 \ = \  2 J_0 - \frac{2}{m_b} \, J_1 \,.
\eeq
For the hadronic matrix elements we use the definitions of the
form factor $f_0$ and $f_+$ as
in \cite{Beneke:2000wa},
\beq
  \langle \pi(p')|\bar q \, \gamma^\mu \, b|B(p)\rangle
   &=& f_+(q^2) \left[p^\mu + p'{}^\mu - \frac{m_B^2}{q^2} \, q^\mu\right]
      +f_0(q^2) \, \frac{m_B^2}{q^2} \, q^\mu
\\[0.3em]
 \langle \pi(p')|\bar q  \, b|B(p)\rangle
   &=& \frac{m_B^2}{m_b} \, f_0(q^2) \,.
\eeq
Here we have neglected the pion mass, and the momentum transfer
is given by $q^2 \simeq m_B (m_B - n_+ p')$.
To leading power in $1/m_b$ we then have
\beq
 m_B \, f_0 &=& \langle \pi| J_0 | B \rangle - \frac{1}{n_+p'}
       \,  \langle \pi| J_1 | B \rangle \, , \\[0.4em]
 (n_+ p') \, f_+ + m_B \, f_0 &=&
 2 \langle \pi| J_0 | B \rangle - \frac{2}{m_B}
       \,  \langle \pi| J_1 | B \rangle \,.
\eeq
With this, the form-factor ratio is given by
\beq
  f_0/f_+ &=& \frac{n_+ p'}{m_B}
   \left( 1 - \frac{2 q^2}{m_B^3} \,  \frac{\langle \pi | J_1 | B \rangle}
                   {\langle \pi | J_0 | B \rangle} + {\cal O}(\alpha_s^2)
  \right) \,.
\eeq
Comparing with Eq.~(62) in \cite{Beneke:2000wa} we identify
\beq
  \frac{\alpha_s C_F}{4\pi} \, \Delta F_\pi &=&
  - \frac{2}{m_B^2} \,  \langle \pi | J_1 | B \rangle  + \ldots
\label{DeltaF}
\eeq
where
\beq
  \Delta F_\pi &=& \frac{8 \pi^2 f_B f_\pi}{3 m_B} \,
   \langle \omega^{-1} \rangle_+^B \, \langle \bar u^{-1} \rangle_\pi
\label{DeltaFres}
\eeq
parametrizes the factorizable form-factor contribution in terms of
the first inverse moments of the $B$\/-meson and pion LCDA.
(Notice, that the definition of the soft form factor $\xi_\pi$
 in \cite{Beneke:2000wa} differs from the one used in \cite{Beneke:2003pa}
and in this work;
but that difference is irrelevant for the form factor {\em ratio}\/ to
the considered order.)

\clearpage

\section*{Erratum}

There is a calculational error in formula (\ref{b12}) 
concerning one of the singly logarithmic terms. The corrected result can be found in
\cite{DeFazio:2007XX}, formula (2.26).

With that, the results in Section~\ref{cancel} 
can be extended, leading now to a perfect cancellation of
scale dependence between hard vertex corrections,
hard-collinear diagrams in SCET and the evolution of $\phi_B^-(\omega)$
derived in \cite{Bell:2008er} (within the Wandzura-Wiczek approximation), 
as it is explictly shown in chapter 2.2.1.\ of \cite{DeFazio:2007XX}.
An update of the numerical analysis can also be found in \cite{DeFazio:2007XX}.

\end{document}